\documentclass[aps,prb,amsmath,amssymb,twocolumn,superscriptaddress,showpacs,floatfix]{revtex4-1}

\usepackage{graphicx}% Include figure files
\usepackage{dcolumn}% Align table columns on decimal point
\usepackage{bm}% bold math

\usepackage{amsmath}
\usepackage{amssymb}

\DeclareMathOperator{\Qg}{\mathit Q_{\mathrm g}}
\DeclareMathOperator{\Sg}{\mathit S_{\mathrm g}}
\DeclareMathOperator{\Cg0}{\mathit C^0_{\mathrm g}}

\DeclareMathOperator{\Cs}{\mathit C_{\Sigma}}
\DeclareMathOperator{\Vg}{\mathit V_{\mathrm g}}
\DeclareMathOperator{\Fm}{\mathit F_{\mathrm m}}
\DeclareMathOperator{\alp}{\alpha_{\mathrm P}}
\DeclareMathOperator{\bp}{\beta_{\mathrm P}}

\begin{document}

% Use the \preprint command to place your local institutional report
% number in the upper righthand corner of the title page in preprint mode.
% Multiple \preprint commands are allowed.
% Use the 'preprintnumbers' class option to override journal defaults
% to display numbers if necessary
%\preprint{}

%Title of paper
\title{Conductance of a SET with a retarded dielectric layer in the gate capacitor}

\author{O.~G.~Udalov}
\affiliation{Department of Physics and Astronomy, California State University Northridge, Northridge, CA 91330, USA}
\affiliation{Institute for Physics of Microstructures, Russian Academy of Science, Nizhny Novgorod, 603950, Russia}
\author{N.~M.~Chtchelkatchev}
\affiliation{L.D. Landau Institute for Theoretical Physics, Russian Academy of Sciences,117940 Moscow, Russia}
\affiliation{Department of Theoretical Physics, Moscow Institute of Physics and Technology, Moscow 141700, Russia}
\affiliation{Institute for High Pressure Physics, Russian Academy of Science, Troitsk 142190, Russia}
\author{S.~A.~Fedorov}
\affiliation{Department of Theoretical Physics, Moscow Institute of Physics and Technology, 141700 Moscow, Russia}
\author{I.~S.~Beloborodov}
\affiliation{Department of Physics and Astronomy, California State University Northridge, Northridge, CA 91330, USA}

\date{\today}


\begin{abstract} We study conductance of a single electron transistor (SET) with a
ferroelectric (or dielectric) layer placed in the gate capacitor. We assume that ferroelectric
(FE) has a retarded response with arbitrary relaxation time. We show that in the case of ``fast''
but still retarded response of the FE (dielectric) layer an additional contribution
to the Coulomb blockade effect appears leading to the suppression of the
SET conductance. We take into account fluctuations of the FE (dielectric)
polarization using Monte-Carlo simulations. For ``fast'' FE these fluctuations
partially suppress the additional Coulomb blockade effect. Using Monte-Carlo
simulations we study the transition from ``fast'' to ``slow'' FE.
For high temperatures the peak value of the SET conductance is almost independent
of the FE relaxation time. For temperatures close to the
FE Curie temperature the conductance peak value non-monotonically depends on
the FE relaxation time. A maximum appears when the FE relaxation time is of
the order of the SET discharging time. Below the Curie point
the conductance peak value decreases with increasing the FE relaxation time.
The conductance shows the hysteresis behavior for
any FE relaxation time at temperatures below the FE transition point.
We show that conductance hysteresis is robust against FE internal fluctuations.
\end{abstract}

%\maketitle must follow title, authors, abstract, \pacs, and \keywords
\maketitle

\section{introduction}

\begin{figure}
\includegraphics[width=1\columnwidth]{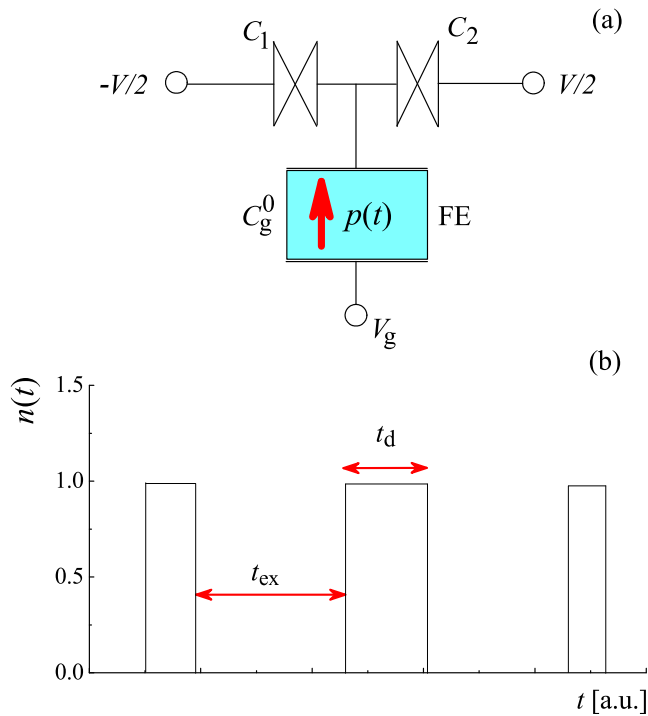}
\caption{(Color online) (a) Circuit diagram of SET coupled to \textbf{a} FE layer. $C_{1,2}$ are the tunnel junctions capacitance, $C_{\mathrm{g}}^0$ is the geometrical gate capacitance, $p(t)$ is the time-dependent polarization of the FE layer inside the gate capacitor, $V$ and $V_\mathrm g$ are the bias and the gate voltages, respectively. (b) Number of excess electrons on the SET metal island $n$ vs. time $t$. SET has the following characteristic times: the electron lifetime on the island (grain discharging time) $t_{\mathrm d}$ and the excitation time $t_{\mathrm{ex}}$.} \label{Fig:fig1}
\end{figure}

Single electron devices such as single electron transistors (SET), single electron (and Cooper pair) boxes, chains of metallic islands and etc., are currently in the focus of scientific interest due to their deep and complex physics and numerous promising applications~[\onlinecite{Giazotto2015, Kouwenhoven2015, You2015, Pekola2015, Oda2015, Khondaker2015, Hakonen2015, Esposito2013, Pekola2014, Armour2013}]. One of the most interesting aspects of these systems is their dynamics. Due to charge quantization even DC current in a metallic chain of islands leads to a complicated dynamics and to the appearance of voltage oscillations similar to those existing in Josephson junctions~[\onlinecite{Serdyukova1989}, \onlinecite{Likharev1986}]. A great progress in fabrication of nanoscale metallic circuits allows creating of single electron devices coupled to nanomechanical resonators. In addition, SETs with moving
metallic island exist. In these systems electrical and mechanical
degrees of freedom are coupled leading to a complex dynamics of the whole device. SET with a mobile metal
particle shows the ``shuttling'' effect~[\onlinecite{Jonson1998,MacKinnon2002, Jonson2002, Jauho2003, Bruder2004}]. Numerous groups studied SET with mobile gate electrode which has its own dynamics~[\onlinecite{Zhang2004, Armour2004, Armour2013, Armour2007, Bruder2006, Brandes2010}]. Even a weak electro-mechanical coupling strongly influences the SET conductance promising very sensitive mechanical sensor. A strong electro-mechanical coupling causes bistability and hysteresis effects which are useful for memory applications.

Recently, a SET coupled to a ferroelectric (FE) layer was proposed and studied~[\onlinecite{Beloborodov2014},\onlinecite{Beloborodov2014_1}]. The FE layer can be placed either in the leads-island capacitors or in the gate capacitor. SET shows high sensitivity of conductivity to the FE dielectric properties. This allows to consider SET as an effective way to study properties of nanosized FE materials. FE hysteresis in combination with its dynamics suggests that ``turnstile'' effect should exist in a SET with FE. This effect is the basis for a precise current standard~[\onlinecite{Averin2014}]. 

When FE layer is placed in the tunneling junctions it serves as tunnel barriers for electrons, but the polarization
of the FE layers have its own dynamics in contrast to the usual SET where the insulator polarization
follows the electric field in the capacitors.

FE placed in the gate capacitor of
SET (see Fig.~\ref{Fig:fig1}(a)) also has its own dynamics. In this case the SET has
much in common with SET having a mobile gate where
movement of the electrode causes change of the gate induced charge.
Variations of the FE polarization influence the SET in the similar way.
However, there are several important differences between these two systems.
Nanomechanical resonators are usually studied as linear oscillators. In contrast,
the FE layer shows the non-linear response since the electric fields in the SET capacitors can exceed the FE saturation field~[\onlinecite{Beloborodov2014},\onlinecite{Beloborodov2014_1}]. Another important difference is related to the fact
that the characteristic time scales of FE materials can be essentially lower than those of
the nanomechanical systems and can be even comparable with characteristic times of SET.
These characteristic timescales of FE layer and the SET determine the transport properties of the whole system.

Different kinds of FE materials show different dynamics. Shift type FEs are described
by the second order (in time) differential equation~[\onlinecite{Levan1983}]. These materials behave similar
to oscillators with a certain resonant frequency $\omega_0^{\mathrm{FE}}$ (in the linear regime)
and a damping time $t_\mathrm{FE}$. Order-disorder type FEs obey the first order differential equation.
These materials are described by the damping time $t_\mathrm{FE}$ only.

The shortest timescale for SET is $t_c \sim \hbar/E_\mathrm c$, where $E_\mathrm c$ is the
charging energy. For a metallic grain of few nm this time
is of order $10^{-14}$ s. An intermediate timescale is the electron lifetime
on the grain (grain discharging time), $t_{\mathrm d}\sim R\Cs$, where $R$ and $\Cs$ are the resistance and the
total capacitance of the system, respectively. This time depends on the geometry of the system
and usually is in the range $10^{-12}$-$10^{-8}$ s.
The longest time $t_{\mathrm{ex}}$ is the time between electron jumps to the grain.
In general this time depends on the gate voltage. For strong Coulomb blockade
it is exponentially large,
$t_{\mathrm{ex}}\sim R\Cs e^{E_\mathrm c/T}$, while for
weak Coulomb blockade, due to applied gate voltage, it is comparable
to the electron lifetime on the grain, $t_{\mathrm d}$.

The limit, $t_\mathrm{FE}\ll t_{\mathrm{c}}$, corresponds to the ``classical'' theory of SET.
In this case the polarization of insulators follows the electric field in the SET capacitors.
When the grain charge changes from $n$ to $n\pm1$ it is implicitly assumed that the
polarization of insulators instantly changes from some equilibrium value $p^{\mathrm{eq}}|_{n}$ to
different equilibrium value $p^{\mathrm{eq}}|_{n\pm 1}$. Thus, the ``classical'' theory of SET considers
transitions between states $(n,p^{\mathrm{eq}}|_{n})$ and $(n\pm1,p^{\mathrm{eq}}|_{n\pm1})$.

The limit, $t_\mathrm{FE}\gg t_{\mathrm{ex}}$, was studied in
Refs.~[\onlinecite{Beloborodov2014},\onlinecite{Beloborodov2014_1}] using the mean field theory.
In this case the polarization of FE layers ``feels'' only the
average electric field produced in the SET capacitors. The
coupling of SET with slow dynamical system causes the hysteresis phenomena
even in the absence of FE polarization hysteresis. In Ref.~[\onlinecite{Beloborodov2014_2}] the SET with ``slow'' dielectric in the gate capacitor was considered.

The region in between the limits discussed above can be divided into
three subregions with the following boundaries: 1) $t_\mathrm{FE}\approx t_{\mathrm{c}}$; 2) $t_\mathrm{FE}\approx t_{\mathrm{d}}$;
and 3) $t_\mathrm{FE}\approx t_{\mathrm{ex}}$. The first region requires consideration of a single tunneling event
by taking into account the time dispersion of dielectric response. This case is beyond
the orthodox theory. Next two regimes can be treated within the orthodox theory
if polarization time dynamics is taken into account.
In the present paper we consider the SET with dielectric or FE layer inside the gate capacitor
with relaxation time larger than $t_{\mathrm{c}}$ ($t_\mathrm{FE}\gg t_\mathrm c$) but with
arbitrary ratio of $t_\mathrm{FE}$ and $t_\mathrm{d,ex}$.

Two different methods are usually used to study dynamical effects in SET.
The first approach calculates the probabilities of all possible states of the system
using master equation~[\onlinecite{Armour2004},\onlinecite{Odintsov1995}]. These probabilities are time-dependent in general. This method is mostly used for investigation of SETs coupled to a mechanical nano resonator or with a mobile metal island. The mechanical subsystem can be treated classically~[\onlinecite{Zhang2004, Bruder2004, Jonson2002, Bruder2006, Brandes2010}] or quantum-mechanically~[\onlinecite{Goan2004,Armour2007,MacKinnon2002, Jauho2003,Hastings2004}].
The second approach simulates the time evolution of the whole system directly
using the Monte-Carlo method~[\onlinecite{Serdyukova1989},\onlinecite{Kosina1997}]. This approach provides the complete information about SET behavior
and allows to analyze time evolution of the system parameters. Currently a single electron counting is possible and time dependence
of the metal island charge can be extracted from an experiment~[\onlinecite{Delsing2005}]. Therefore
the results of Monte-Carlo simulations can be compared to a real experimental data.

In the present paper we use the Monte-Carlo simulations along with analytical consideration
to study SET coupled to a FE layer behaving classically. In particular, we study the influence
of the retarded response of the FE layer
in the gate capacitor on the Coulomb blockade effects.

The paper is organized as follows. In Sec.~\ref{Sec:Model} we describe the model of the SET with FE (dielectric) layer
in the gate capacitor. In Sec.~\ref{Sec:Results} we discuss our results: we consider analytically and numerically the SET conductance in the intermediate region $t_\mathrm{c}\ll t_\mathrm{FE}\ll t_\mathrm{d}$. We discuss the evolution of the SET conductance with increasing
the FE relaxation time from  $t_\mathrm{FE}\ll t_\mathrm{d}$ to $t_\mathrm{FE}\gg t_\mathrm{ex}$. We investigate
the influence of the FE (dielectric) internal noise on the SET conductance.

\section{Model}\label{Sec:Model}

We consider the SET shown in Fig.~\ref{Fig:fig1}(a). Left and right electrodes (source and drain) are connected
to a metallic island via tunnel junctions. The source and the drain are biased with voltage $\pm V/2$. For simplicity,
we assume that all junctions have the same capacitances $C_1=C_2=C/2$ and resistances $R_1=R_2=R$. An insulator
with an instant response is placed inside the tunnel junctions. The polarization of the insulator instantly reacts to
the electric field, $p_{1,2}=\chi_{\mathrm{jun}}(\pm V/2-\phi)/d$, where $\phi$ is the
grain potential, $d$ is the junctions thickness and $\chi_\mathrm{jun}$ is the dielectric susceptibility of the insulator in the junctions. The
polarizability of the tunnel barriers is incorporated into capacitances $C_{1,2}$. A gate electrode is capacitively
coupled to the metallic island.  A voltage $V_\mathrm g$ is applied to the electrode.
The FE (dielectric) with retarded response is placed between the island and the gate.
The characteristic time scale of the material, $t_\mathrm{FE}$, is larger than $t_\mathrm{c}$.
We assume that the polarization $p(t)$ of the FE (dielectric) layer in the gate capacitor is uniform and time dependent.
We use the notation $C_{\mathrm g}^0 =\Sg/(4\pi d_{\mathrm g})$ for geometrical capacitance of the gate-island capacitor,
where $\Sg$ and $d_{\mathrm g}$ are the gate capacitor area and thickness, respectively.

Below we will distinguish two different temperatures: 1) $T_\mathrm{FE}$ is the FE layer temperature and
2) $T_\mathrm e$ is the leads and the island temperature. In general, these temperatures can be different
on one hand due to the current flowing through the source-island-drain circuit which heats the island and the leads~[\onlinecite{Luukanen2006,Vasenko1994,Pekola2007}] and on the other hand different temperatures can be created
artificially using local heating/cooling techniques.~[\onlinecite{Chapuis2006,King2011,Khurgin2007,Chen2011}].

\subsection{Free energy of SET with FE (dielectric) layer in the gate capacitor}

Equations governing the behavior of SET with a FE (dielectric) layer can be
derived using the free energy of the system. The free energy
increment can be expressed via charges $q_i$ and potentials $\phi_i$
of all the metallic electrodes $\delta F=\delta R=\sum\phi_i \delta q_i$.
Here the charges are considered as independent variables. In a SET the three
metallic electrodes are biased with a voltage source and the island potential should be found self-consistently.
Following the standard approach~[\onlinecite{landauVol8}] we introduce the thermodynamic potential where
the island charge, the potentials of the leads and the gate potential are considered as independent variables $\delta F_{\mathrm m}=\delta F-\delta(V_\mathrm g\Qg)-\delta(V_1 q_1)-\delta(V_2 q_2)= \phi\delta q  -\Qg\delta V_\mathrm g-q_1\delta V_1 -q_2 \delta V_2$, where $\Qg$ is the charge
of the gate electrode, $q_{1,2}$ and $V_{1,2}$ are the charges and the potentials of the leads, $q$ and $\phi$ are the metal island
charge and potential, respectively. Below we use the notation  $F_\mathrm m$ for free energy.

At zero bias voltage the increment of this mixed thermodynamic potential
has the form $\delta F_{\mathrm m}=\phi\delta q-\Qg\delta V_\mathrm g$.
Integration of $\delta \Fm$ over the $\delta q$ and $\delta V_\mathrm g$ provides
the total free energy. Using electrostatic consideration for the charge
at the gate electrode we find $\Qg=(C\Cg0 V_\mathrm g-q\Cg0-p\Sg C)/\Cs$, where
$\Cs=C+\Cg0$ is the total capacitance of the system.
The grain potential is given by $\phi=(\Cg0 V_\mathrm g+q-p\Sg)/\Cs$. Thus,
\begin{equation}\label{Eq_FreeEnSET}
\Fm=F_0+\frac{q^2}{2\Cs}+\frac{q\Cg0 V_\mathrm g}{\Cs}-\frac{p\Sg q}{\Cs}+\frac{p\Sg C V_\mathrm g}{\Cs}-\frac{C\Cg0 V^2_\mathrm g}{2\Cs},
\end{equation}
where $F_0$ is the system free energy at zero gate voltage and zero grain charge
\begin{equation}\label{Eq_FreeEnFE}
F_0=(\alp p^2/2+\bp p^4/4+p^2\Sg/(2\Cs d_{\mathrm g}))\Sg d_{\mathrm g}.
\end{equation}
The first two terms in Eq.~(\ref{Eq_FreeEnFE})
are the usual contributions to the free energy describing
the dielectric materials and FEs
close to the  paraelectric-ferroelectric phase transition~[\onlinecite{Levan1983}]. Here
$\alp$ and $\bp$ are the phenomenological constants. For dielectric materials both constants
are positive. For FEs the constant $\alp$ linearly depends
on temperature and crosses zero at the FE phase transition point.
The last term in Eq.~(\ref{Eq_FreeEnFE})
is due to the non-zero electric field $E_0=-4\pi \Cg0 p/\Cs$ acting
on the FE layer at $q=0$ and $V_\mathrm g=0$.

At finite bias voltage, $V$, the free energy $F_\mathrm m$ needs to be modified:
$F_\mathrm m$ gets an additional contribution $\pm eV/2$ when an electron is added to the grain.

For linear instant relation between the polarization $p$ and the
electric field, $p(t)=\chi^{\mathrm i}_0(\phi(t)-V_\mathrm g)/d_{\mathrm g}$, we obtain the
free energy of the SET with the properly renormalized
capacitances $C_{\mathrm g}=\Cg0 \epsilon_0$ and
$\tilde C_{\Sigma}=C+C_{\mathrm g}$, where $\epsilon_0=1+4\pi\chi^{\mathrm i}_0$ is the effective
dielectric constant of the layer in the gate capacitor.

\subsection{Ferroelectric dynamics \label{sec:FEeq}}

Varying the free energy in Eq.~(\ref{Eq_FreeEnSET}) with respect to the polarization
$p$ we find the equation governing $p$. Below we consider
FEs (dielectrics) of the order-disorder type. These materials are described using the first order
time dependent equation~[\onlinecite{Levan1983}]
\begin{equation}\label{Eq:OrderDis}
\gamma\dot p=-\frac{\partial \Fm}{\partial p}+\tilde\Gamma_{\mathrm L},
\end{equation}
where $\gamma$ is the relaxation constant related to the relaxation time,
$t_\mathrm{FE}=\gamma\chi_0$, with $\chi_0=\alpha^{-1}_\mathrm P$ being
the linear response of the polarization to the external electric field.
$\tilde\Gamma_{\mathrm L}$ describes an intrinsic noise of the FE layer.
Substituting the free energy, $F_\mathrm m$, into Eq.~(\ref{Eq:OrderDis}) we obtain
the following non-linear equation for polarization
\begin{equation}\label{Eq:ShiftFE_NL}
\gamma\dot p+\alpha_{\mathrm P}p+\beta_{\mathrm P}p^3=\frac{q-C V_\mathrm g-p \Sg}{\Cs d_{\mathrm g}}+\tilde\Gamma_{\mathrm L}.
\end{equation}
We introduce the paraelectric-ferroelectric Curie temperature at which the coefficient in front of the linear term is zero,
$\chi^{-1}=\alpha_\mathrm P+4\pi\Cg0/\Cs=\tilde \alpha_\mathrm P(T_{\mathrm{FE}}-T_\mathrm C)=0$. We call the material inside
the gate capacitor as a FE material if the Curie temperature is positive $T_\mathrm C>0$.
For $T_\mathrm{FE}>T_\mathrm C$, Eq.~(\ref{Eq:ShiftFE_NL}) has only one solution, while for
$T_\mathrm{FE} < T_\mathrm C$ the spontaneous polarization appears leading to multiple solutions of the equation.

Another Curie temperature $T^0_\mathrm C$ can be introduced using
the relation $\chi_0^{-1}=\tilde\alpha^0_\mathrm P(T_\mathrm{FE}-T^0_\mathrm C)=0$.
It corresponds to the same FE but confined by the shorted metal electrodes. One can see
that $T_\mathrm C<T^0_\mathrm C$. When a FE is placed in between the
shorted metal electrodes the potential difference and the electric field
are fixed across the FE layer. In contrast, in SET problem the potential
of the metal island should be found self-consistently.
Therefore the equation governing the FE layer gives the lower Curie temperature.
Physically this reflects the fact that the depolarizing electric field inside the FE layer is not fully
screened by the metal island in the SET. This complicates the appearance of the spontaneous polarization.

We notice that $\chi_0$ is the susceptibility of the FE (dielectric) layer
with respect to the external electric field, while $\chi$ is the susceptibility
with respect to the external charges $q$ and $Q_0=-\Cg0\Vg$. Below
we will refer to $Q_0$ as the gate charge. However, $Q_0$ is not the real charge at the gate electrode and it differs
from $Q_\mathrm g$ which was introduced above.
For $T_{\mathrm{FE}}>T_\mathrm C$ two susceptibilities $\chi$ and $\chi_0$ are related
as follows $\chi^{-1}=\chi^{-1}_0+4\pi\Cg0/\Cs$. For
FE (dielectric) materials with instant response the constant
$\chi_0$ defines the dielectric permittivity, $\chi^{\mathrm i}_0=\chi_0$ and $\epsilon_0=1+4\pi\chi_0$.

We regard the material in the gate capacitor as a dielectric if the
quantity, $\alpha_\mathrm P+4\pi\Cg0/\Cs > 0$, is positive for
any temperature. In general, we can consider the dielectric material as the
material with the negative Curie temperature, $T_\mathrm C<0$.

In our Monte-Carlo simulations we solve the non-linear Eq.~(\ref{Eq:ShiftFE_NL}) directly.
For analytical consideration at temperatures $T_\mathrm{FE}>T^0_\mathrm C$
we linearize Eq.~(\ref{Eq:ShiftFE_NL})
\begin{equation}\label{Eq:ShiftFE1}
\gamma\dot q_{\mathrm p}+\chi^{-1}q_{\mathrm p}=\frac{4\pi \Cg0(q-C V_\mathrm g)}{\Cs}+\Gamma_{\mathrm L}.
\end{equation}
Here we introduce the charge $q_{\mathrm p}=p\Sg$ and $\Gamma_{\mathrm L}=S_\mathrm g \tilde \Gamma_\mathrm L$.

We mention that below $T^0_\mathrm C$ the susceptibility $\chi_0$ becomes negative.
However, the FE layer is still in the paraelectric phase for temperatures
$T_\mathrm C<T_\mathrm{FE}<T^0_\mathrm C$. We show that in this temperature region
Eq.~(\ref{Eq:ShiftFE1}) is not valid since it produces divergent solutions
for $q_\mathrm p$ and $q$. The nonlinear term $\beta_\mathrm P p^3$ should
be taken into account to restrict variations of $q_\mathrm p$ and $q$.

\subsection{Langevin forces}

Internal fluctuations of the FE layer occur due to the interaction of FE (dielectric) polarization with all other degrees of freedom in the FE layer. This interaction has two components. The ``regular'' component
leads to the appearance of polarization relaxation. This term can be written in the form $\gamma \dot q_{\mathrm p}$ even for nonlinear systems~[\onlinecite{Zwanzig1973}]. Another noise component
is responsible for fluctuations of the FE layer polarization. It appears in the RHS of Eq.~(\ref{Eq:ShiftFE_NL}) as a random Langevin force $\tilde\Gamma_{\mathrm L}$. If polarization varies slowly in time in comparison to characteristic time of internal FE processes the correlation function of the Langevin forces can be chosen as $C_{\mathrm{L}}=\langle \tilde\Gamma_{\mathrm L}(t) \tilde\Gamma_{\mathrm L}(t')\rangle=2\gamma k_\mathrm B T_\mathrm{FE} \delta(t-t')/(S_\mathrm g d_\mathrm g)$. The relation between the relaxation constant $\gamma$ and the Langevin forces dispersion is valid
even for nonlinear systems~[\onlinecite{Zwanzig1973}].
The dispersion of the polarization fluctuations in the linear response regime
($\beta_\mathrm P\ll\alpha_\mathrm P p^2$) can be written using the FE layer susceptibility as follows
$D_{\mathrm p}=\langle (\Delta p)^2\rangle=\chi k_{\mathrm B} T_{\mathrm{FE}}/(\Sg d_{\mathrm g})$. Below
we will use this expression for estimates. For strong nonlinear effects ($\beta_\mathrm P\sim\alpha_\mathrm P p^2$)
the polarization dispersion is given by more complicated expression~[\onlinecite{Zwanzig1971}].

\subsection{SET dynamics \label{sec:FEeq}}

The probability of electron hop to/from the metallic island is given by the expression
\begin{equation}\label{Eq:Prob1}
G^{\pm}_{1,2}=\frac{1}{e^2R}\frac{\Delta F^{\pm}_{1,2}}{e^{\Delta F^{\pm}_{1,2}/T_\mathrm e}-1},
\end{equation}
where
\begin{equation}\label{Eq:Prob2}
\begin{split}
\Delta F^{\pm}_1=2E_\mathrm c\left(\pm\left(n+\frac{\Cg0 V_\mathrm g}{e}+\frac{\Cs V}{2e}-\frac{q_{\mathrm p}}{e}\right)+1/2\right),\\
\Delta F^{\pm}_2=2E_\mathrm c\left(\pm\left(n+\frac{\Cg0 V_\mathrm g}{e}-\frac{\Cs V}{2e}-\frac{q_{\mathrm p}}{e}\right)+1/2\right).
\end{split}
\end{equation}
Subscript ($1,2$) denotes the electrode from which (or to which) an electron jumps. Superscript $+$ describes
electron hopping from a certain lead to the particle. Superscript $-$ describes
electron hopping from the island to a certain lead. $n=q/e$ is the island population.
$E_\mathrm c=e^2/(2\Cs)$ is the bare charging energy.

Important assumption made when calculating $\Delta F$ is that
polarization $p$ does not change during the electron hop since
$t_\mathrm{FE}\gg t_\mathrm c$. Equation~(\ref{Eq:Prob2}) describes the
processes ($n, p(t)$)$\to$($n\pm1, p(t)$). This is in contrast to
the ``classical'' SET situation where polarization $p$ adjusts
during the hop and the following processes occur ($n, p^{\mathrm{eq}}|_n$)$\to$($n\pm1, p^{\mathrm{eq}}|_{n\pm1}$).

Two characteristic times can be introduced using Eq.~(\ref{Eq:Prob1}): 1) the
excitation time $t_{\mathrm{ex}}$ corresponding to the island charging event at zero gate voltage, $V_\mathrm g=0$.
This event is suppressed due to Coulomb blockade effect,
$t_{\mathrm{ex}}=G^{+}_1|_{n=0,V_{\mathrm g}=0,q_\mathrm p=0,V=0}\sim R\Cs e^{E_\mathrm c/T_\mathrm e}$;
2) the relaxation time $t_{\mathrm d}$ corresponding to the discharging of the charged island.
This time is much shorter, $t_{\mathrm{ex}}\gg t_{\mathrm d}$, since there is
no exponential factor in this case, $t_{\mathrm d}=G^{-}_1|_{n=1,V_{\mathrm g}=0,q_\mathrm p=0,V=0}\sim R\Cs$.
We mention that excitation time depends strongly on the gate voltage $V_\mathrm g$. And for
certain voltages both times are comparable.

The island population is defined by the random electron jumps from and to the island
\begin{equation}\label{Eq:GrEl1}
\dot n(t)=\sum_{i}Z_i\delta(t-t_i).
\end{equation}
Here $Z_i$ has three possible values $\pm 1$ and $0$.
The probability for electron to
hop per unit time is defined by Eq.~(\ref{Eq:Prob1}). It depends on the
bias voltage  $V$, the gate voltage $V_\mathrm g$, the number of electrons
on the island $n(t)$ and the FE layer induced charge $q_{\mathrm p}(t)$. The last two quantities
depend on time leading to time dependent hopping probabilities.
Thus, one has to solve self-consistent equations for the FE layer and the SET.

We use numerical Monte-Carlo simulations~[\onlinecite{Serdyukova1989}] to solve
coupled SET and FE equations. Equation~(\ref{Eq:GrEl1}) can be written in the discrete form
\begin{equation}\label{Eq:GrElDiscr1}
\begin{split}
&n(t_i)=n(t_{i-1})+Z(n(t_{i-1}),q_\mathrm p(t_{i-1}))dt,\\
&Z=\left\{\begin{array}{l} ~1,~~ x<G^{+}_1+G^{+}_2,\\~0,~~ G^{+}_1+G^{+}_2<x<1-(G^{-}_1+G^{-}_2), \\ -1,~~ x>1-(G^{-}_1+G^{-}_2), \end{array}\right.
\end{split}
\end{equation}
where $x$ is the random value uniformly distributed between $0$ and $1$, $dt=t_i-t_{i-1}$.
The time interval $dt$ is much shorter than all characteristic times
in the problem, $dt\ll t_{\mathrm d},~t_\mathrm{FE}$.

The average electric current through the SET can be calculated as follows
\begin{equation}\label{Eq:Curr}
\begin{split}
&I=\frac{e}{T_\mathrm m}\sum_{i}\zeta(t_i),\\
&\zeta=\left\{\begin{array}{l} ~~1,~~ Z(t_i)>0 \land y<G^{+}_1/(G^{+}_1+G^{+}_2),\\-1,~~Z(t_i)<0 \land y<G^{-}_1/(G^{-}_1+G^{-}_2), \\
~~0,~~Z(t_i)=0,\end{array}\right.
\end{split}
\end{equation}
where  $y$ is the random number
uniformly distributed between $0$ and $1$; $T_\mathrm m$ is the measurement time interval.

\section{Conductance of SET with retarded FE (dielectric) layer}\label{Sec:Results}

We investigate the behavior of SET conductance as a function of
gate voltage, $V_\mathrm g$ for different ratios of the FE (dielectric) relaxation time,
$t_\mathrm{FE}$, and SET characteristic times $t_{\mathrm{ex}}$ and $t_\mathrm d$.

\subsection{Suppression of Coulomb blockade due to internal noise of FE}

The conductance peaks as a function of gate charge $Q_0$ have a finite width
at finite temperatures. The width of these peaks is defined as
$\Delta Q_0^{\mathrm L}\approx e k_{\mathrm B} T_\mathrm e/E_\mathrm c$.
The polarization of the FE (dielectric) layer enters the SET equations as an additional
gate capacitor charge. Therefore the FE (dielectric) polarization fluctuations can be considered as
fluctuations of the gate charge $Q_0$.
These fluctuations average the SET conductance over
the region $\Delta Q_0^\mathrm p\sim\sqrt{D_{\mathrm p}} \sim\sqrt{T_\mathrm{FE}}$.
Due to the square root dependence of $\Delta Q_0^\mathrm p$ on temperature,
the internal fluctuations of the FE polarization produce
much stronger effect on the conductance than the finite temperature of the leads.
If temperature of the FE layer $T_\mathrm{FE}$ equals to the source and drain electrodes
temperatures, $T_\mathrm e$, then the FE fluctuations are negligible for temperatures
$k_{\mathrm B}T_\mathrm{FE} \ll 2\pi\chi\Cg0 E_\mathrm c/\Cs$. Also, FE fluctuations can be neglected
for temperatures, $T_\mathrm{FE}\ll T_\mathrm e$.

If $\Delta Q_0^\mathrm p$ exceeds the half of an electron charge
the Coulomb blockade effect is smeared and the conductance peaks as a function
of the gate voltage disappear. Thus, the general condition for observation of
charge quantization effects relates the polarization dispersion and the
electron charge, $S_\mathrm g \sqrt{D_\mathrm p}\ll e/2$.

Below we show that FE fluctuations can increase the SET conductance in some cases.

\subsection{SET with ``instant'' FE (dielectric) in the gate capacitor}

We call FEs (dielectrics) with $t_\mathrm{FE}\ll t_{\mathrm{c}}$ as FEs (dielectrics) with
instant response, for brevity we call them ``instant'' FEs (dielectrics).
In contrast, we call FEs ``retarded'' if $t_\mathrm{FE}\gg t_{\mathrm{c}}$.
A SET with ``instant'' dielectrics was studied in many papers using the orthodox theory.
In this case the conductance is a periodic function of the gate charge $Q_0$. The
conductance maxima are located at points $Q_0=e(l+1/2)/\epsilon_0$, where $l$ is an integer number.
The peak value of the SET conductance in this case is $\sigma_0=1/(4R)$.
The conductance is suppressed between peaks due to the Coulomb blockade effect.

If SET is coupled to a FE with relaxation time $t_\mathrm{FE}\approx t_\mathrm{c}$,
the SET can not be studied within the orthodox theory. This case requires
the calculation of tunneling matrix elements of electrons interacting
with FE (dielectric) polarization varying on time scale $t_\mathrm{c}$.
Tunneling in the presence of time dispersion was considered
in the past~[\onlinecite{Urbina1990,Stiles1990,devoret1992single}]. It was shown that
time dispersion, caused by the dissipation in the tunnel junction circuit, leads to
the decrease of the junction conductance and to the appearance of an
effective Coulomb blockade effect.

\subsection{SET with ``fast'' FE (dielectric) in the gate capacitor ($t_\mathrm{FE}\ll t_{\mathrm d}$)}

In the case of ``fast'' FE (dielectric), $t_\mathrm c\ll t_\mathrm{FE}\ll t_{\mathrm d}$, each electron jump
to the grain or out of grain occurs when polarization is in it's equilibrium state
at a given grain population $n$. However, during the jump the FE (dielectric) polarization
preserves it's value since $t_\mathrm{FE}\gg t_{\mathrm{c}}$, meaning that the state
of the FE polarization is not equilibrium after the jump. We define the state of the SET
at each moment with the pair ($n(t),p(t)$). The following transitions correspond to the
events of grain charging and discharging ($n,p|^{\mathrm{eq}}_{n}$)$\to$($n\pm 1,p|^{\mathrm{eq}}_{n}$), where $p|^{\mathrm{eq}}_n$ stands for equilibrium polarization at given $n$ and $Q_0$
\begin{equation}\label{Eq:FE_NL_eq}
\chi^{-1}p^{\mathrm{eq}}|_n+\beta_{\mathrm P}(p^{\mathrm{eq}}|_n)^3=\frac{en+CQ_0/\Cg0}{\Cs d_{\mathrm g}}.
\end{equation}
We mention that for SET with ``instant'' insulators different kind of transitions occur ($n,p|^{\mathrm{eq}}_{n}$)$\to$($n\pm 1,p|^{\mathrm{eq}}_{n\pm1}$).

Consider the free energy difference in Eq.~(\ref{Eq:Prob2}) as a function of
island population $n$ at $V=0$. The stable population of the island is approximately defined
by the conditions
\begin{equation}\label{Eq:EquilCond}
\left\{\begin{array}{l}
\Delta F^+(n+1,p^{\mathrm{eq}}|_{n+1})>\Delta F^-(n+1,p^{\mathrm{eq}}|_{n+1}),\\
\Delta F^+(n,p^{\mathrm{eq}}|_{n})<\Delta F^-(n,p^{\mathrm{eq}}|_{n}).
\end{array}
\right.
\end{equation}
We find approximately the ``equilibrium'' value of $n$ and polarization
$p$ at a given value of $Q_0$ using the following equation
\begin{equation}\label{Eq:Equilibr}
\Delta F^+(n^{\mathrm{eq}},p^{\mathrm{eq}}|_{n^{\mathrm{eq}}})=\Delta F^-(n^{\mathrm{eq}},p^{\mathrm{eq}}|_{n^{\mathrm{eq}}}).
\end{equation}

\begin{figure}
\includegraphics[width=0.9\columnwidth]{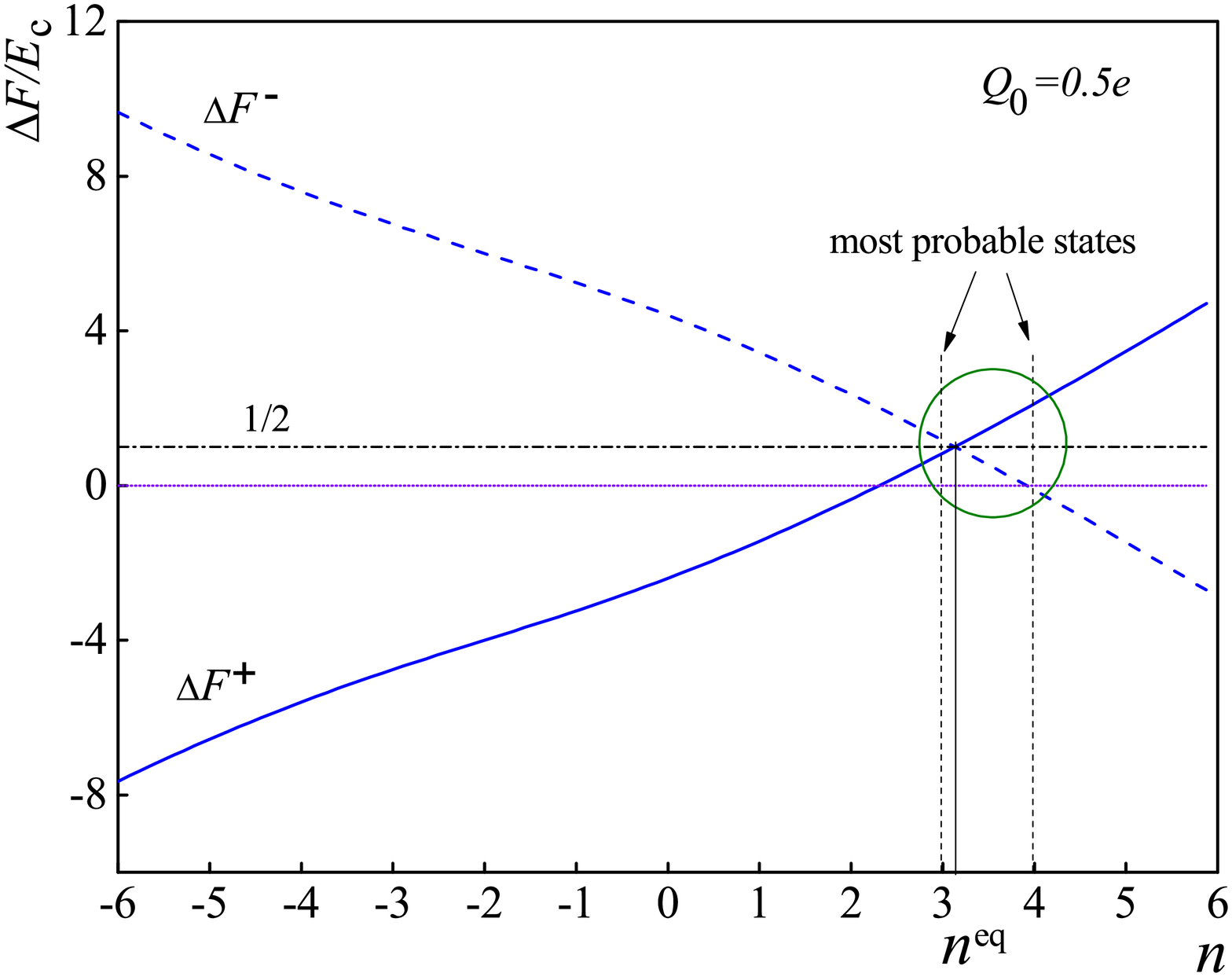}
\caption{(Color online) Free energy difference, Eq.~(\ref{Eq:Prob2}), at $V=0$ and $Q_0=0.5$e
vs. $n$. Solid line shows $\Delta F^+(n)$, dashed line shows $\Delta F^-(n)$.
Intersection of $\Delta F^+$ and $\Delta F^-$ shows the most
probable state of the SET at a given $Q_0$.} \label{Fig:FreeEnDiff1}
\end{figure}

Here $n^{\mathrm{eq}}$ stands for approximately the most probable population of the island,
it may have a non-integer value. Equation~(\ref{Eq:Equilibr}) can be rewritten in the form
\begin{equation}\label{Eq:Equilibr1}
en^{\mathrm{eq}}-Q_0-q^{\mathrm{eq}}_\mathrm p|_{n^{\mathrm{eq}}}=0,
\end{equation}
where $q^{\mathrm{eq}}_\mathrm p|_{n^{\mathrm{eq}}}=\Sg p^{\mathrm{eq}}|_{n^{\mathrm{eq}}}$. Using Eq.~(\ref{Eq:FE_NL_eq}) we can relate the ``equilibrium'' polarization and the gate charge $Q_0$
\begin{equation}\label{Eq:Equilibr2}
\alpha_\mathrm P (q^{\mathrm{eq}}_\mathrm p|_{n^{\mathrm{eq}}})+\tilde\beta_\mathrm P (q^{\mathrm{eq}}_\mathrm p|_{n^{\mathrm{eq}}})^3=4\pi Q_0,
\end{equation}
where $\tilde \beta_\mathrm P=\beta_\mathrm P/ S_\mathrm g^2 $.
Equation~(\ref{Eq:Equilibr2}) has a unique solution for temperatures
$T_\mathrm{FE}>T^0_\mathrm C$. For $T_\mathrm{FE}<T^0_\mathrm C$ it has three different solutions.
Introducing $p^{\mathrm{eq}}|_{n^{\mathrm{eq}}}$ into Eq.~(\ref{Eq:Equilibr1}) we find the ``equilibrium'' population $n^{\mathrm{eq}}$. At a given $Q_0$ the whole system state most probably will
be in the vicinity of $n^{\mathrm{eq}}$. To find approximately the SET conductance we
take into account only two states $n_1$ and $n_2$ ($n_1<n_2$) nearest to $n^{\mathrm{eq}}$. Using the expression
\begin{equation}\label{Eq:Cond1}
\sigma=\frac{e}{V}\frac{\tilde G^+_2 \tilde G^-_1- \tilde G^+_1 \tilde G^-_1}{\tilde G^+_1+\tilde G^+_2+\tilde G^-_1+\tilde G^-_2}
\end{equation}
we find the SET conductance. We use the following notations in Eq.~(\ref{Eq:Cond1}):
$\tilde G^+_{1,2}=G^+_{1,2}|_{n=n_1}$, $\tilde G^-_{1,2}=G^-_{1,2}|_{n=n_2}$. Below
we consider the behavior of SET conductance at different temperatures.

\subsubsection{$T_\mathrm{FE}>T^0_\mathrm C$.}

\begin{figure}
\includegraphics[width=1\columnwidth]{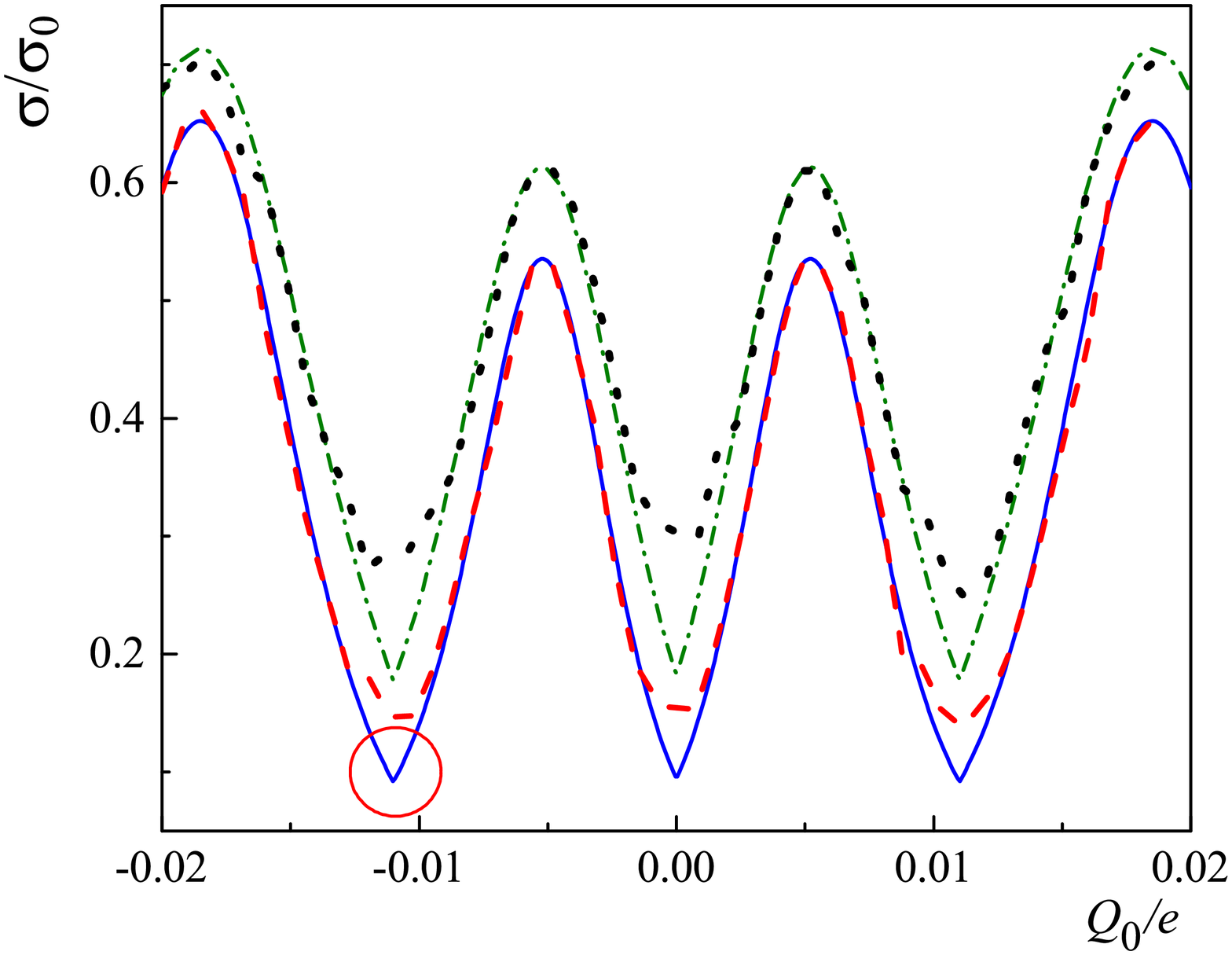}
\caption{(Color online) SET conductance vs. the gate charge $Q_0$ for temperatures
$T_\mathrm{FE}>T^0_\mathrm C$. Solid blue line is
calculated using Eq.~(\ref{Eq:Cond1}) at electron temperature $T_e=0.2E_\mathrm c$.
Dashed red line and dotted black line are obtained using MC simulations. Dashed line
is calculated in the absence of FE internal fluctuations. Dotted black line is
calculated in the presence of fluctuations. Green dash dotted line is
calculated with Eq.~(\ref{Eq:Cond1}) at $T_e=0.26 E_\mathrm c$. Red circle shows
the typical region where Eq.~(\ref{Eq:Cond1}) is not valid.} \label{Fig:CondVg0}
\end{figure}

In this temperature region the parameter $\alpha_\mathrm P$ is positive
and Eq.~(\ref{Eq:Equilibr2}) has only one solution leading to one ``stable'' state for a whole system. Figure~\ref{Fig:FreeEnDiff1} shows the free energy difference as a function of the island population $n$ for
$\Cg0=0.2\Cs$, $\alpha_\mathrm P=1.5$, $e^2\tilde\beta_\mathrm P=0.13$ and $Q_0=0.5e$. Inequalities~(\ref{Eq:EquilCond})
are satisfied for $n=3$ only. Equation~(\ref{Eq:Equilibr2}) gives the ``equilibrium'' population $n^{\mathrm{eq}}$
slightly higher than 3. The most probable states are $n=3$ and $n=4$.
Figure~\ref{Fig:CondVg0} shows the SET conductance (Eq.~(\ref{Eq:Cond1})) as a function of the gate charge $Q_0$
for parameters $\Cg0=0.005\Cs$, $\alpha_\mathrm P=0.12$, $e^2\tilde\beta_\mathrm P=0.0013$, and $T_\mathrm e=0.2$ $E_\mathrm c$. These parameters correspond to TTF-CA FE at $T_\mathrm{FE}=120$ K~[\onlinecite{Tokura2012}]. This FE has rather small relaxation time in the ps region~[\onlinecite{Okamoto2013}], which can be comparable, smaller or larger than discharging time of SET.
The conductance appears as series of peaks. The period and the height of the peaks are not constant due to nonlinear term in Eq.~(\ref{Eq:ShiftFE_NL}).
The red circle shows the typical region where Eq.~(\ref{Eq:Cond1}) is not valid. In this region at least three states are involved into electron transport, whereas Eq.~(\ref{Eq:Cond1}) accounts for only two states. The
dashed line shows the conductance calculated using numerical Monte-Carlo (MC) simulations. We use $\gamma=0.05 R\Cs$. The numerical and analytical results coincide in the vicinity of the conductance peaks,
where Eq.~(\ref{Eq:Cond1}) works well. Close to the conductance deeps the numerical result is more
accurate since all possible states are taken into account in the simulations.
The conductance peaks value is smaller than $\sigma_0$ for ``fast'' FE. Below we
show that this is a consequence of an effective additional Coulomb blockade appearing due to the retarded response of the
FE layer.

It is important that Eq.~(\ref{Eq:Cond1}) does not take into account internal fluctuations of the
FE polarization. This approximation is valid, for example, for temperatures
$T_\mathrm{FE}\ll T_{\mathrm e}$. To fit blue curve in Fig.~\ref{Fig:CondVg0}
with MC simulations we turned off the Langevin forces in Eq.~(\ref{Eq:ShiftFE_NL}). The result is the red dashed line.

The dotted black line in Fig.~\ref{Fig:CondVg0} shows the result of MC modelling for
$T_\mathrm{FE}=T_\mathrm e$ in the presence of FE internal fluctuations.
One can see that the FE fluctuations are important for electron transport.
The internal fluctuations suppress the additional Coulomb blockade increasing
the conductance. For comparison we show the conductance calculated using
Eq.~(\ref{Eq:Cond1}) for the same SET parameters but with higher electron
temperature, $T_e=0.26$ $E_\mathrm c$ (dash dotted green line).
\begin{figure}
\includegraphics[width=1\columnwidth]{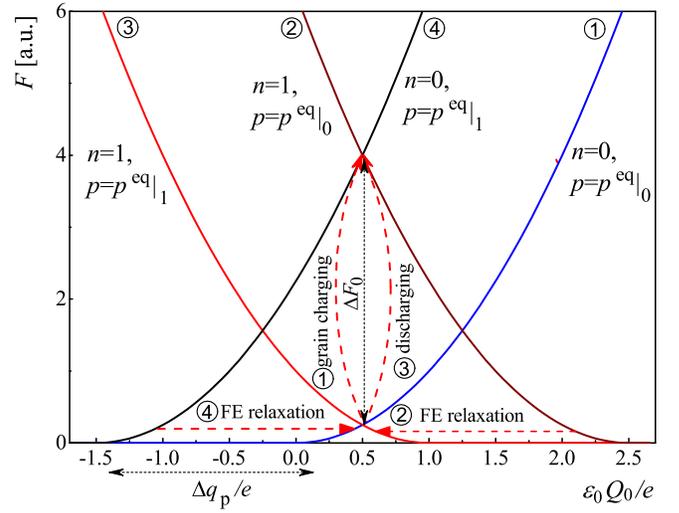}
\caption{(Color online) Free energy as a function of gate charge $Q_0$.
Blue branch (1) corresponds to the state ($0,p^{\mathrm{eq}}|_{0}$), red branch (3) corresponds to the state ($1,p^{\mathrm{eq}}|_{1}$), brown branch (2) shows the free energy for the state ($1,p^{\mathrm{eq}}|_{0}$)
and black branch (4) shows the state ($0,p^{\mathrm{eq}}|_{1}$).
At $\epsilon_0 Q_0=e/2$ in the SET with ``instant'' insulators the grain charging and discharging events correspond to the transition from blue to the red branch. Energy gap in this case is zero leading to large
conductance. For SET with ``retarded'' FE another kind of transitions occurs.
The island charging event (1) corresponds to the transition from
blue branch (1) to the brown branch (2). After the transition the FE polarization
relaxes to a new equilibrium state (process depicted by the horizontal arrow (2)).
The grain discharging event (3) corresponds to the transition from
red (3) to the black (4) branch. Finally, the FE polarization relaxes
again (process (4)).
$\Delta F_0=4\pi\chi\Cg0E_\mathrm c/ C_{\mathrm \Sigma}$, $\Delta q_\mathrm p=(4\pi \chi \Cg0/\Cs)e.$} \label{Fig:Expl}
\end{figure}

For small enough parameter $\beta_\mathrm P$ we can neglect non-linear effects and use the linearized equation for the FE polarization, Eq.~(\ref{Eq:ShiftFE1}). In this case the SET conductance is a periodic function of parameter
$Q_0$ with the period $\Delta Q_0=e/\epsilon_0$. Consider the conductance in the vicinity of the first peak position $Q_0=-e/(2\epsilon_0)$. Figure~\ref{Fig:Expl} shows different processes contributing to the electron transport
of the SET with ``retarded'' FE layer. This figure shows a number of branches of the SET
free energy as a function of the gate charge $Q_0$. These branches correspond to the island population $n=0,1$ and the
FE polarization $p^{\mathrm{eq}}|_{0,1}$. For SET with ``retarded'' FE the transitions (charging and discharging events) occur between the states with non-equilibrium polarization of the FE (dielectric) layer. This leads to the
appearance of a non-zero energy gap (see Fig.~\ref{Fig:Expl}) for any gate voltage.
Such a gap can be considered as an additional effective Coulomb blockade appearing due to ``retarded'' FE (dielectric) response.

At $\epsilon_0 Q_0=e/2$  the charging event ($0,p|^{\mathrm{eq}}_{0}$)$\to$($1,p|^{\mathrm{eq}}_{0}$) (process (1) in Fig.~\ref{Fig:Expl}) requires
an additional energy $\Delta F_0=4\pi\chi\Cg0E_\mathrm c/ C_{\mathrm \Sigma}$.
The FE layer relaxes to it's equilibrium state $p|^{\mathrm{eq}}_{1}$ (the process (2) in Fig.~\ref{Fig:Expl})
after the grain charging event.
The discharging process (3) ($1,p|^{\mathrm{eq}}_{1}$)$\to$($0,p|^{\mathrm{eq}}_{1}$) is also inelastic.
It requires the same energy $\Delta F_0$.
This energy gap suppresses the conductance of the SET with ``retarded'' FE layer.
For SET with ``instant''  FE (dielectric) layer the charging and discharging transitions
occur directly from branch (1) to branch (3) and do not require any
energy at $\epsilon_0 Q_0=e/2$ leading to a higher conductance.

\begin{figure}
\includegraphics[width=1\columnwidth]{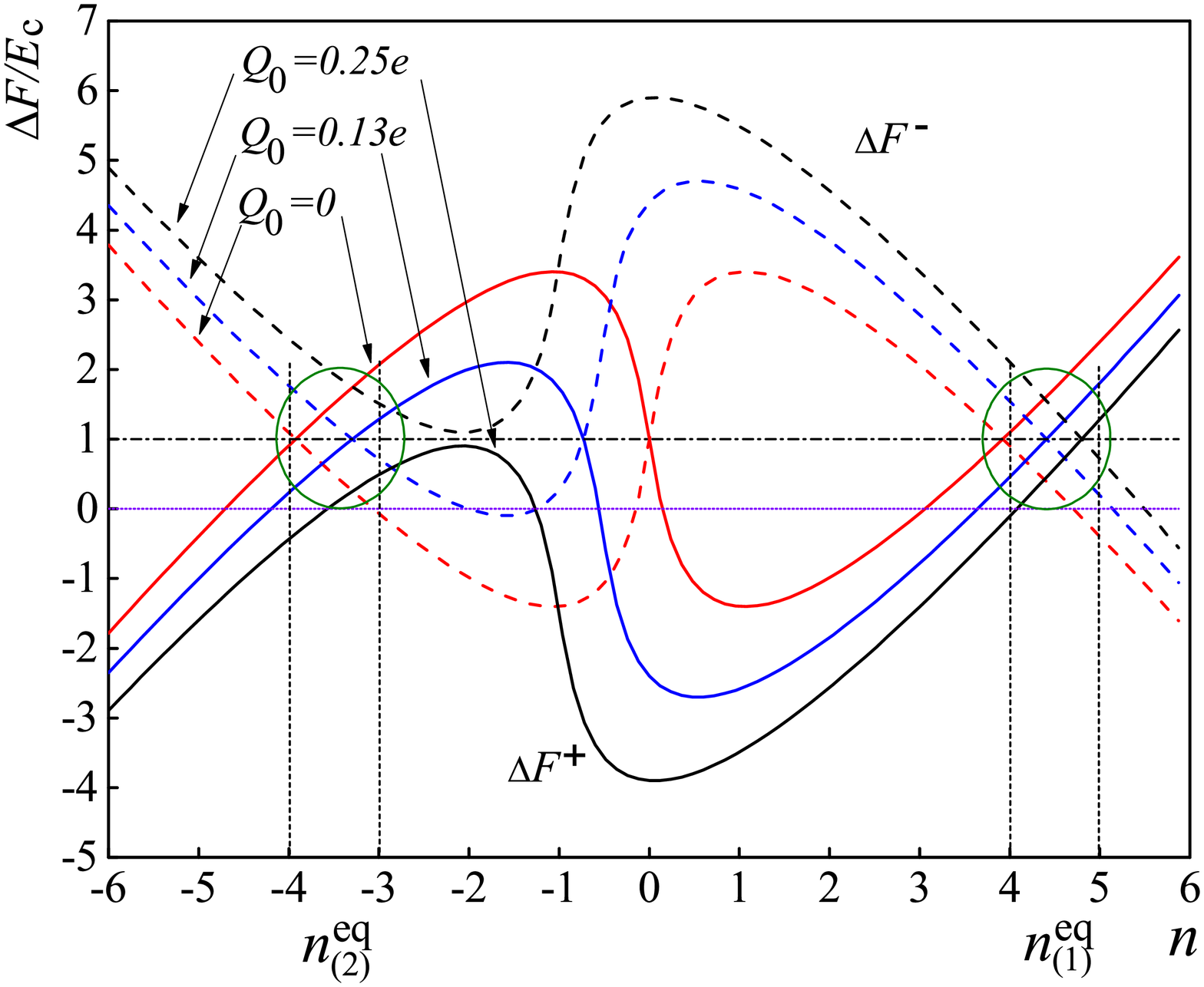}
\caption{(Color online) Free energy difference, Eq.~(\ref{Eq:Prob2}), at $V=0$ vs. $n$.
Solid lines show $\Delta F^+(n,p^{\mathrm{eq}}|_n)$, dashed lines show $\Delta F^-(n,p^{\mathrm{eq}}|_n)$. Red lines correspond to $Q_0=0$, blue lines show $\Delta F^{\pm}$ at $Q_0=0.13e$  and black lines show
the case of $Q_0=0.25e$. Negative $\chi_0$ provides the negative slope
of $\Delta F^{+}$ and positive slope of $\Delta F^{-}$ in the vicinity of $n=0$.} \label{Fig:FreEn1}
\end{figure}

For temperatures $T_\mathrm e\ll E_\mathrm c (4\pi\chi\Cg0/ C_{\mathrm \Sigma})$ the maximum
conductance value is
\begin{equation}\label{Eq:MaxCond}
\sigma_{\mathrm{max}}\approx\frac{1}{R}\frac{2\pi\chi\Cg0}{C_{\mathrm \Sigma}}\frac{E_\mathrm c}{T_\mathrm e} e^{-\frac{\Delta F_0}{T_\mathrm e}}.
\end{equation}
The magnitude of conductance peak increases exponentially with temperature, however it
stays smaller than the ``classical'' value of SET conductance $\sigma_0$.
We mention that for SET with ``instant'' insulators the magnitude of conductance peak
is temperature independent.

\subsubsection{$T_\mathrm C< T_\mathrm{FE}<T^0_\mathrm C$}

In this temperature region the susceptibility $\chi_0$ is negative and Eq.~(\ref{Eq:Equilibr2}) has three different solutions
meaning that there are more than one stable state in the whole SET system. Figure~\ref{Fig:FreEn1} shows
$\Delta F^{\pm}(n,p^{\mathrm{eq}}|_n)$ for different $Q_0$ for the following SET parameters $\alpha_\mathrm P=-2$, $\chi^{-1}=0.513$, $e^2\tilde\beta_\mathrm P=0.13$, $\Cg0=0.2 \Cs$. 

Here the dependencies are non-monotonic in contrast to the case $T_\mathrm{FE}>T^0_\mathrm C$.
Consider the blue curves corresponding to $Q_0=0.13e$. One can see that the state with zero island
population is not stable.
Two stable states, shown by the green circles, occur in the vicinity of $n^{\mathrm{eq}}_{(1)}\approx 4.5 e$ and $n^{\mathrm{eq}}_{(2)}\approx -3.5 e$. Depending on the initial system
state both ``equilibrium'' populations can be realized. In this case one can expect the hysteresis behavior of
the SET conductance as a function of the gate voltage.
Figure~\ref{Fig:FreEn1} shows that for high enough gate voltage (black lines) only
one stable state remains meaning that conductance hysteresis exists
only around $Q_0=0$. This is in contrast to the case of SET with ``slow'' FE considered in Refs.~[\onlinecite{Beloborodov2014},\onlinecite{Beloborodov2014_1}], where hysteresis appears in the vicinity of each conductance peak.
The nature of hysteresis in the case of ``slow'' FE in the gate capacitor is very
different from the case of ``fast'' FE (dielectric). For ``fast'' FE the hysteresis
is related to the instability ($\chi_0<0$) of the FE dielectric properties, while for
``slow'' FE the hysteresis occurs even in the absence of FE instability due
to the interaction of the fast SET system with the slow FE layer.

Note that in the region $T_\mathrm C< T_\mathrm{FE}<T^0_\mathrm C$ the nonlinear term in Eq.~(\ref{Eq:ShiftFE_NL}) for the FE polarization can not be neglected since the population $n(t)$ and the polarization $p(t)$ will be divergent functions.
\begin{figure}
\includegraphics[width=0.95\columnwidth]{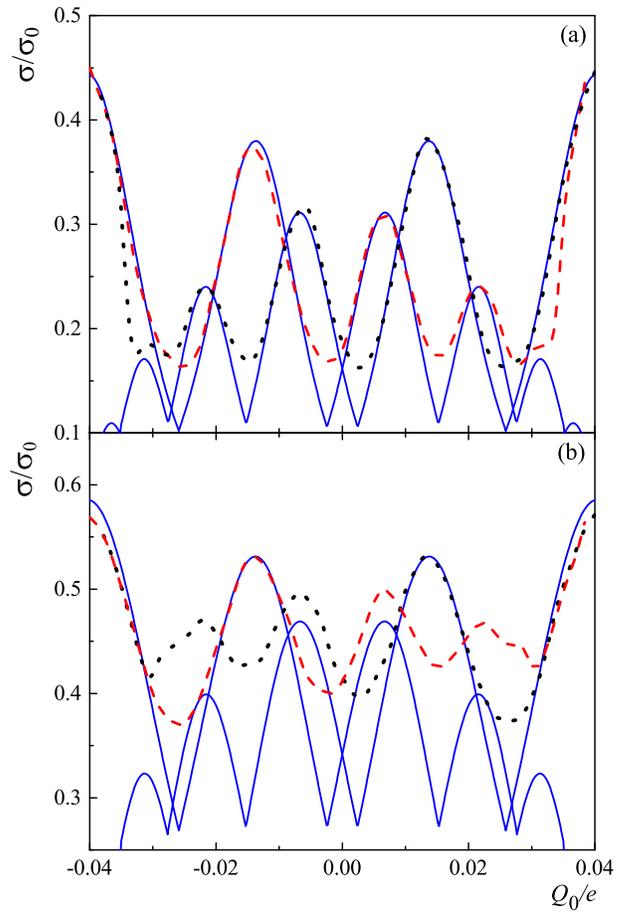}
\caption{(Color online) Conductance of SET as a function of gate charge $Q_0$ for temperatures
$T_\mathrm C<T_\mathrm{FE}<T^0_\mathrm C$. Solid blue lines are calculated
using Eq.~(\ref{Eq:Cond1}).
In the vicinity of $Q_0=0$ there are two branches of the conductance due to the SET instability.
Dashed red line and dotted black line are calculated using MC simulations.
Dashed line is obtained for increasing parameter $Q_0$ during simulations, while the
dotted line is obtained for decreasing $Q_0$. Panel (a) shows results obtained
using Eq.~(\ref{Eq:Cond1}) at $T_\mathrm e=0.2 E_\mathrm c$ and results of MC simulations
at $T_\mathrm e=0.2 E_\mathrm c$ in the absence of FE internal fluctuations.
Panel (b) shows results obtained using Eq.~(\ref{Eq:Cond1})
at $T_\mathrm e=0.3 E_\mathrm c$ and results of MC simulations in the presence of
FE internal fluctuations and at $T_\mathrm e=0.2 E_\mathrm c$.} \label{Fig:CondVg1}
\end{figure}

Figure~\ref{Fig:CondVg1} shows the SET conductance as a function of the gate charge $Q_0$ for the following parameters: $\alpha_\mathrm P=-0.12$, $e^2\tilde\beta_\mathrm P=0.0013$, $\Cg0=0.02 \Cs$. These parameters correspond to TTF-CA FE at $T_\mathrm{FE}\approx 50$ K~[\onlinecite{Tokura2012}].  Solid lines demonstrate the result obtained using Eq.~(\ref{Eq:Cond1}) at $T_\mathrm e=0.2E_\mathrm c$ (panel (a)) and $T_\mathrm{e}=0.3 E_\mathrm c$ (panel (b)). One can see that there are two branches of the conductance as a function of the gate voltage. Dashed and dotted lines demonstrate the conductance obtained using MC simulations in the absence
of FE internal fluctuations (panel (a)) and in the presence of fluctuations (panel (b)). To obtain the
red dashed lines we increase $Q_0$ from $-0.5e$ to $+0.5e$. The initial polarization and the
island population at each $Q_0$ were taken from the last simulation point at previous $Q_0$.
Dotted lines correspond to the case of decreasing $Q_0$. MC simulations confirm the existence of
the conductance hysteresis. The analytical and numerical curves agree well with each over.
Part of the conductance branches, for example between $Q_0=0.1e$ and $Q_0=0.25e$, obtained using
Eq.~(\ref{Eq:Cond1}) is not realized in numerical simulations within the chosen protocol.
To observe these branches we need to set a special initial conditions in our system which
is hard to realize in an experiment.
\begin{figure}
\includegraphics[width=1\columnwidth]{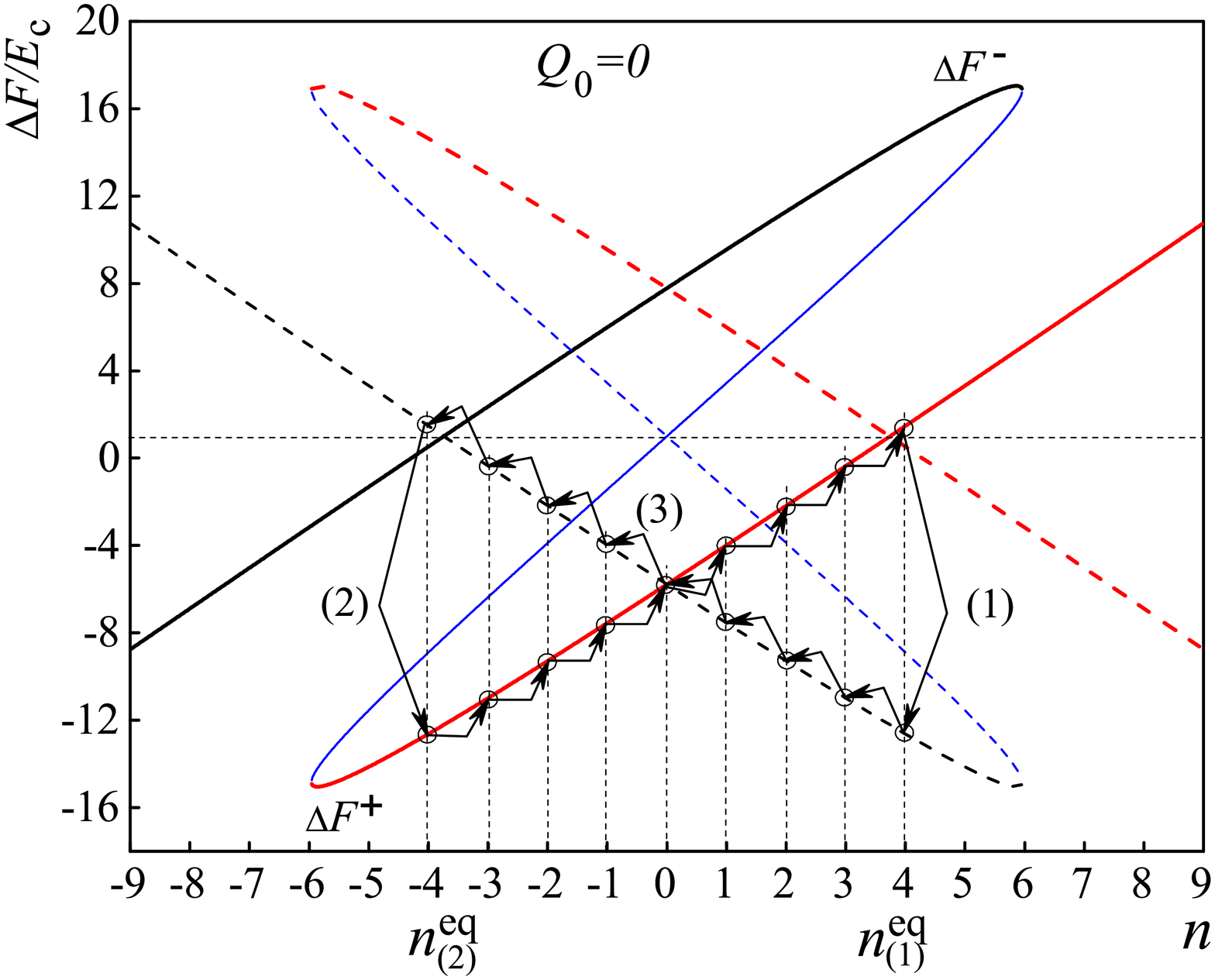}
\caption{(Color online) Free energy difference, Eq.~(\ref{Eq:Prob2}),
at $V=0$ and $Q_0=0$ vs. $n$. Solid lines show $\Delta F^+(n,p^{\mathrm{eq}}_{1,2}|_n)$, dashed lines show $\Delta F^-(n,p^{\mathrm{eq}}_{1,2}|_n)$. Various colors correspond to different FE layer polarization states. Blue lines show the unstable polarization state, $p^{\mathrm{eq}}_3|_n$, which can not be realized in
an experiment. Red line stands for positive polarization, black line corresponds to
negative polarization. Black arrows show possible behavior of the SET. } \label{Fig:FreEn3}
\end{figure}

Panel (b) shows the result of simulations in the presence of FE internal fluctuations
($T_\mathrm{FE}=T_\mathrm e$). Once can see that fluctuations suppress the Coulomb blockade
and increase the conductivity. The peak positions stay the same. The FE fluctuations at
given parameters do not smear the hysteresis.

\subsubsection{$T_\mathrm{FE}<T_\mathrm C$}

In this region the solution of Eq.~(\ref{Eq:ShiftFE_NL}) becomes ambiguous leading
to a more complicated behavior of the SET. Two different equilibrium polarizations
$p^{\mathrm{eq}}_{1,2}|_n$ correspond to a single island population value $n$.
Consider the free energy difference as a function of $n$ at $\alpha_\mathrm P=-7$, $e^2\tilde\beta_\mathrm P=0.5$, $\Cg0=0.1 \Cs$, $Q_0=0$ (Fig.~\ref{Fig:FreEn3}). Quantities $\Delta F^{\pm}$ have two branches depicted by red and black curves. These branches correspond to equilibrium polarization states $p^{\mathrm{eq}}_{1}|_n$ and $p^{\mathrm{eq}}_{2}|_n$. Blue lines show unstable solutions. ``Equilibrium'' population  $n^{\mathrm{eq}}_{(1)}$ is defined
by the intersection of two red lines. In contrast to the case of $T_\mathrm{FE}>T_\mathrm C$, there is an additional
allowed state at $n=n^{\mathrm{eq}}_{(1)}$ corresponding to another FE polarization.
Due to inequalities (\ref{Eq:EquilCond}) this second state is unstable with respect to variations of $n$.
The internal fluctuations of the FE polarization may cause switching of the polarization (process (1) in Fig.~\ref{Fig:FreEn3}) leading to the transition to another ``equilibrium'' population state $n^{\mathrm{eq}}_{(2)}$. We depict a possible dynamical process in the SET in Fig.~\ref{Fig:FreEn3} by the black arrows.
The system switches back and forth between two population states, $n=n^{\mathrm{eq}}_{(1)}$ and $n=n^{\mathrm{eq}}_{(2)}$.

The temperature driven transition between two FE equilibrium polarization states
is defined by the energy barrier between these two states, which
can be estimated as $\Delta W_\mathrm p\approx 2S_\mathrm g d_\mathrm g\alpha^2_\mathrm P/(\beta_\mathrm p)=2\alpha^2_\mathrm P/(4\pi\tilde\beta\Cg0)$. Varying parameters one can make this barrier much higher than $E_\mathrm c$, thus prohibiting the polarization switching. In this case there is no population ``wandering'' and the whole system behaves similarly to the case of $T_\mathrm{FE}>T_\mathrm C$. In particular,
the conductance shows hysteresis as a function of $Q_0$. For parameters used
in Fig.~\ref{Fig:FreEn3}, $\Delta W_\mathrm p>100 E_\mathrm c$.
Therefore processes (1) and (2) are almost prohibited. The hysteresis behavior in
the system with above parameters is rather robust.

The hysteresis can also be smeared through the consecutive
change of $n$. The transition between $n=n^{\mathrm{eq}}_{(1)}$ and $n=n^{\mathrm{eq}}_{(2)}$
through such changes requires at least energy of order $16E_\mathrm c$.
This estimate is obtained using Fig.~\ref{Fig:FreEn3} as a maximum value of
the red dashed curve. Moreover, the SET should go through several
transitions ($n$)$\to$($n-1$) to reach the equilibrium state $n=n^{\mathrm{eq}}_{(1)}$
from $n=n^{\mathrm{eq}}_{(2)}$. Thus, the probability of such a switching between
equilibrium states is negligible.
\begin{figure}
\includegraphics[width=1\columnwidth]{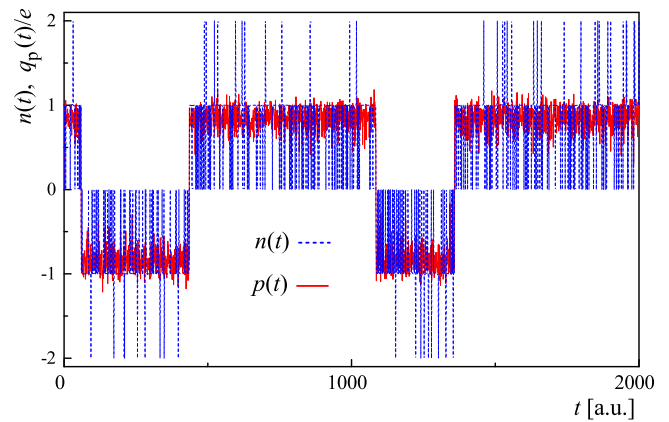}
\caption{(Color online) Island population $n$ (blue dashed line) and FE layer induced charge, $q_\mathrm p/e$,
(red solid line) vs. time $t$ for the following parameters $\alpha_\mathrm P=-0.15$, $e^2\tilde\beta_\mathrm P=0.2$, $\Cg0=0.002\Cs$, $Q_0=0$. The FE layer polarization switches between two equilibrium states due to thermal fluctuations.} \label{Fig:Wandering}
\end{figure}

For barrier $\Delta W_\mathrm p\sim k_\mathrm B T_\mathrm{FE}$ the system switches
between states $n=n^{\mathrm{eq}}_{(1)}$ and $n=n^{\mathrm{eq}}_{(2)}$. This
behavior is shown in Fig.~\ref{Fig:Wandering} where we depict
functions $n(t)$ and $p(t)$ obtained using MC simulation with
the following parameters $\alpha_\mathrm P=-0.15$, $e^2\tilde\beta_\mathrm P=0.2$, $\Cg0=0.002 \Cs$, $Q_0=0$.
Our MC simulations show that in this case there is no hysteresis behavior of the conductance
as function of $Q_0$. This happens because the switching between
different polarization states $p^{\mathrm{eq}}_{1,2}|_n$ means that on the long time
scale the FE layer behaves as a paraelectric layer due to strong fluctuations. Note that such a mechanism of hysteresis
suppression does not exist for temperatures $T_\mathrm C < T_\mathrm{FE}<T^0_\mathrm C$.
In this region the polarization $p$ has always a single value.
The suppression of hysteresis due to consecutive change of $n$ is very
improbable too. Therefore, the conductance hysteresis is rather robust
in the region $T_\mathrm C < T_\mathrm{FE}<T^0_\mathrm C$, however it is not very
stable for $T_\mathrm{FE} < T_\mathrm{C}$.

\subsection{SET with ``slow'' FE (dielectric) in the gate capacitor ($t_\mathrm{FE}\gg t_{\mathrm{ex}}$)}

Characteristic time $t_{\mathrm{ex}}$ is the time between consecutive electron jumps to the grain.
It depends on the gate voltage and it is comparable
with $t_{\mathrm d}$ for weak Coulomb blockade.

The intermediate region, $t_{\mathrm{d}}\ll t_\mathrm{FE}\ll t_{\mathrm{ex}}$, shows similar
results for the conductance peak value as the region $t_\mathrm{FE}\gg t_{\mathrm{ex}}$. However,
the peak shape in the intermediate region is quantitatively different
since changing the gate voltage $V_\mathrm g$ the ratio $t_\mathrm{FE}/ t_{\mathrm{ex}}$ changes too. However, our calculations show that the difference in the shape of the peaks is not qualitative. Therefore it is difficult
to observe it in an experiment. Thus, we consider both the intermediate and the ``slow'' FE regions
simultaneously.

The limit of slow FE was studied before using the mean field theory.
In particular, two different situations were discussed in Refs.~[\onlinecite{Beloborodov2014_1},\onlinecite{Beloborodov2014_2}]. In Ref.~[\onlinecite{Beloborodov2014_2}] the SET with linear dielectric
material in the gate capacitor was discussed.
It was shown that due to coupling of fast SET system with slow dielectric
system the hysteresis of the conductance as a function of $Q_0$ appears at large positive
susceptibility of the dielectric layer ($\chi_0>\chi_0^\mathrm{cr}=T_\mathrm e\Cs/(\pi\Cg0(E_\mathrm c +T))$).
The Ref.~[\onlinecite{Beloborodov2014_1}] studied the SET with strongly
non-linear FE material in the gate capacitor. Similar to the case of ``fast''
FE layer the hysteresis in the vicinity of the point $Q_0=0$ was predicted.
In the present paper we study the conductance behavior moving from ``fast''
to ``slow'' FE (dielectric) in the presence of FE (dielectric) polarization fluctuations.
These fluctuations substantially modify the behavior of SET even in the case of ``slow'' FE.

Following the mean field theory of the SET with ``slow'' and
weak dielectric layer ($0<\chi_0<\chi_0^\mathrm{cr}$) [\onlinecite{Beloborodov2014_1}] in the gate capacitor
the conductance in the vicinity of $Q_0=e/(2\epsilon_0)$ has the form
\begin{equation}\label{Eq:CondPeak}
\sigma^{\mathrm{MF}}/\sigma_0\approx\frac{e\delta Q_0/(\Cs k_\mathrm B T_\mathrm e)}{\sinh(e\delta Q_0/(\Cs k_\mathrm B T_\mathrm e))}\approx 1-\frac{1}{6}\left(\frac{e\delta Q_0}{\Cs k_\mathrm B T_\mathrm e}\right)^2,
\end{equation}
where $\delta Q_0=Q_0-e/(2\epsilon_0)$.
Equation~(\ref{Eq:CondPeak}) is valid for small deviations $\delta Q_{\mathrm 0}$.
The FE fluctuations lead to averaging of the maximum over a finite region of $Q_0$.
This region is proportional to the square root of the FE fluctuations
dispersion, $S_\mathrm g\sqrt{D_{\mathrm p}}\sim\sqrt{T_{\mathrm{FE}}}$. For
equal temperatures $T_{\mathrm{FE}}=T_{\mathrm e}=T$ the averaging gives the following maximum
value for conductance
\begin{equation}\label{Eq:CondPeakAv}
\begin{split}
\sigma_{\mathrm{max}}^{\mathrm{av}}= &\sigma_0 \left(1-\frac{1}{18}\left(\frac{eS_{\mathrm g}\sqrt{D_{\mathrm p}}}{\Cs k_\mathrm B T}\right)^2\right)=\\
=&\sigma_0\left(1-\frac{1}{18}\frac{4\pi \Cg0 e^2\chi}{\Cs^2 k_\mathrm B T}\right).
\end{split}
\end{equation}
Equation~(\ref{Eq:CondPeakAv}) shows that the averaging over fluctuations
leads to the decrease of conductance peak value. Moreover, with decreasing the
temperature the conductance peak value decreases. This unexpected behavior
is related to the fact that fluctuations of FE polarization behaves
as $\sqrt{T}$ leading to the ``strong'' $\sim\sqrt{T}$ free energy difference fluctuations.
Below we will show that FE fluctuations also smear the hysteresis appearance for $\chi_0>\chi_0^\mathrm{cr}$).

\subsection{Transition from ``fast'' to ``slow'' FE (dielectric)}

The analytical consideration is difficult for $t_\mathrm{FE}\sim t_\mathrm d$.
Therefore we use MC simulations to study this region.
\begin{figure}
\includegraphics[width=1\columnwidth]{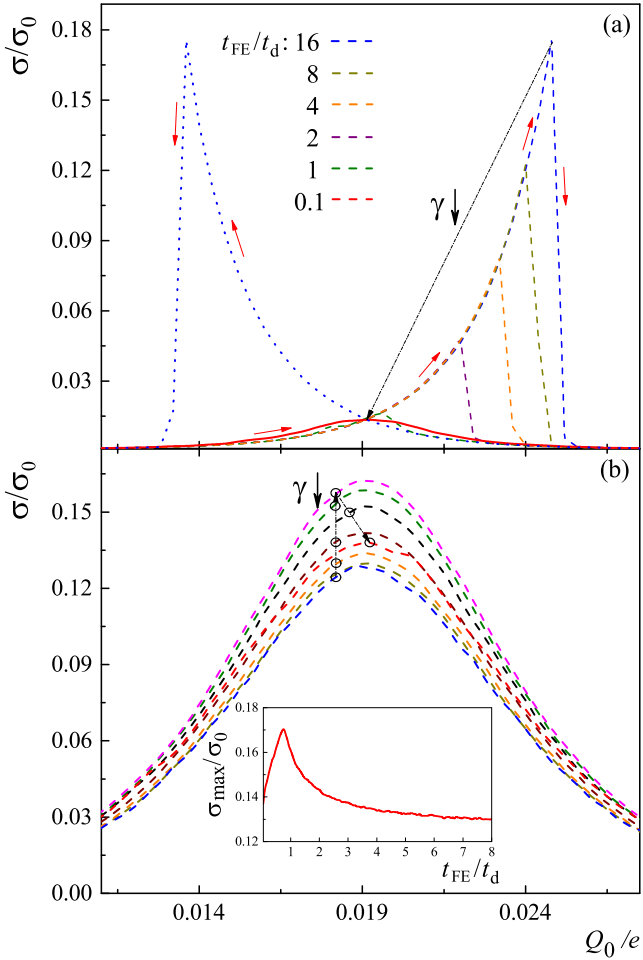}
\caption{(Color online) Conductance of SET as a function of $Q_0$ for different $t_\mathrm{FE}$.
(a) FE fluctuations are absent ($T_\mathrm{FE}\ll T_\mathrm e$). Red arrows show
the path around the hysteresis loop. Both hysteresis branches are shown
for $t_\mathrm{FE}/t_\mathrm d=16$. (b) FE fluctuations are present
($T_\mathrm{FE}=T_\mathrm e$). Inset in the panel (b) shows the dependence of the peak
value of the conductance on $t_\mathrm{FE}$.} \label{Fig:SETgamma1}
\end{figure}
\begin{figure}
\includegraphics[width=1\columnwidth]{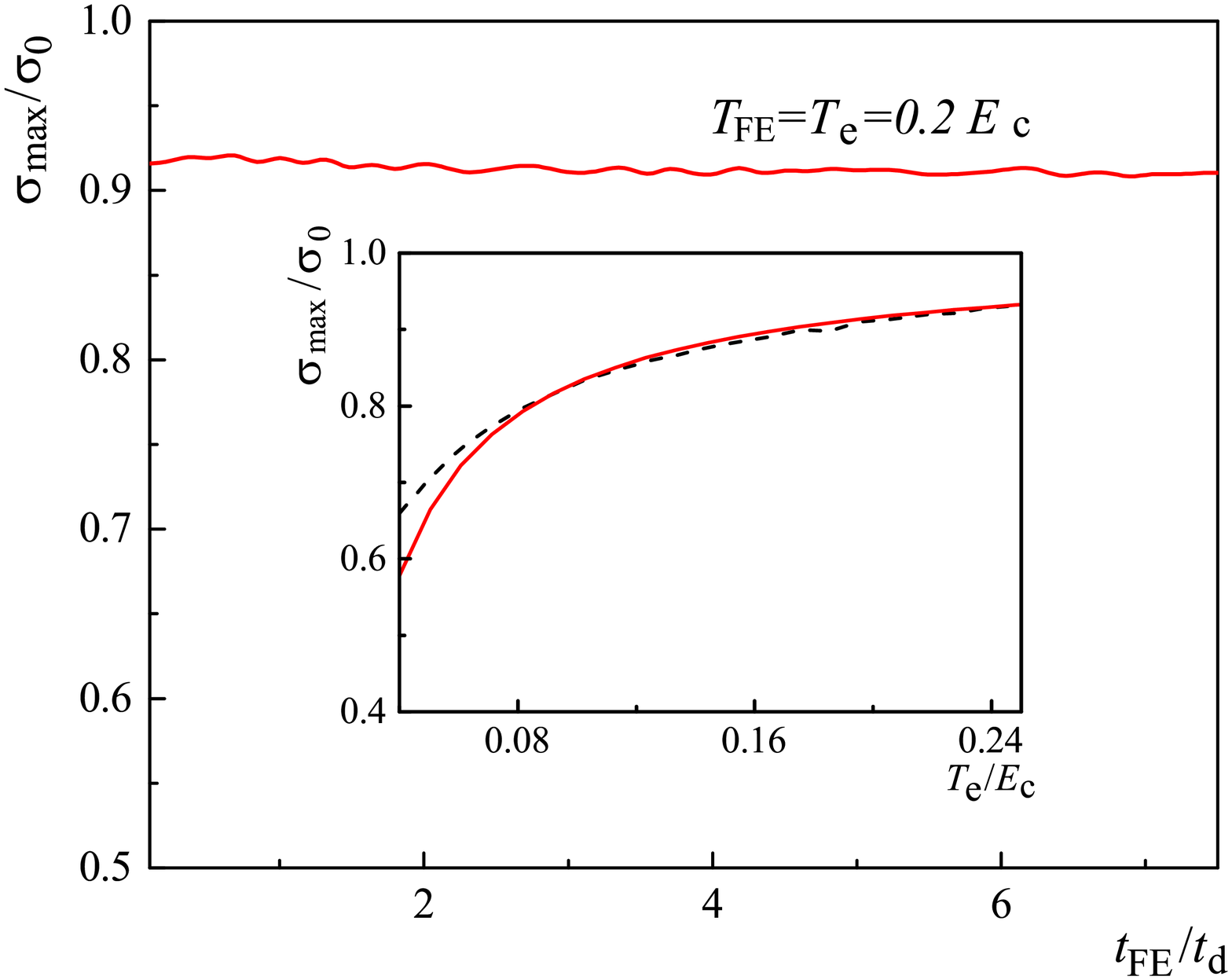}
\caption{(Color online) Conductance peak value vs. $t_\mathrm{FE}$ for small
FE (dielectric) susceptibility, $\chi_0$. The inset shows the conductance peak
value at $t_\mathrm{FE}/t_\mathrm d=8$ as a function of
temperature $T_\mathrm e =T_\mathrm{FE}$. Dashed black line shows the result of MC
simulations. Red solid line corresponds to the modified Eq.~(\ref{Eq:CondPeakAv}). } \label{Fig:SmallSusc}
\end{figure}

\subsubsection{$T_\mathrm{FE}>T^0_\mathrm C$.}

In the case of ``fast'' FE (dielectric) the conductance appears as
series of peaks as a function of the gate charge $Q_0$. Here we consider
the case of linear FE (dielectric) response. In this case the period of peaks
and height is constant and does not depend on $Q_0$. The peak height is defined
by an effective additional Coulomb blockade. In the case of ``slow'' FE (dielectric) the
conductance is also a periodic function of $Q_0$ (see Refs.~[\onlinecite{Beloborodov2014_1},\onlinecite{Beloborodov2014_2}]) with
the same period as in the case of ``fast'' dielectric. However,
the shape of conductance peaks strongly depends on the magnitude of the
dielectric susceptibility, $\chi_0$. For large susceptibility the conductance shows
hysteresis behavior as a function of $Q_0$ while for small susceptibility the hysteresis is absent.
The hysteresis behavior was demonstrated in Refs.~[\onlinecite{Beloborodov2014_1},\onlinecite{Beloborodov2014_2}] using
the mean field theory in the absence of fluctuations of the FE (dielectric) polarization.
Our numerical simulations show that FE fluctuations destroy
the hysteresis for temperatures $T_\mathrm{FE}=T_\mathrm e = T$.
Lowering the FE temperature $T_\mathrm{FE}$ decreases the fluctuations of $Q_0$ but
the hysteresis width also decreases with electronic
temperature $T_\mathrm e$. Therefore to observe the hysteresis in the case of
``slow'' dielectric it is necessary to use the SET with different temperatures
of the dielectric layer and the current conducting circuit, $T_\mathrm{FE}<T_\mathrm e$.

Figure~\ref{Fig:SETgamma1} shows the evolution of
function $\sigma(Q_0)$ with increasing parameter $\gamma$ for
large susceptibility $\chi_0$ for the following parameters:
$\Cg0=0.03 \Cs$, $\alpha_\mathrm P=0.5$, $e^2\tilde\beta_\mathrm P=3\cdot 10^{-4}$, $T_\mathrm e=0.06 E_\mathrm c$.
Panel (a) shows the ideal case in the absence of FE fluctuations and temperatures
$T_\mathrm{FE}\ll T_\mathrm e$.
For large FE relaxation time, $t_\mathrm{FE}\gg t_\mathrm d$, the hysteresis is
very pronounced, see blue dashed and blue dotted lines. Red solid arrows show the
path around the hysteresis loop. The hysteresis width and the maximum value of conductance decrease
with decreasing time $t_\mathrm{FE}$ ($\gamma$). For
$t_\mathrm{FE}\ll t_\mathrm d$ the hysteresis disappears and the conductance is
suppressed due to an additional Coulomb blockade effect.
The situation is very different for finite FE fluctuations
and temperatures $T_\mathrm{FE}=T_\mathrm e$, see panel (b).
In this case the conductance hysteresis is absent for $t_\mathrm{FE}\gg t_\mathrm d$.
The conductance as a function of $Q_0$ has a peak in the
vicinity of $\epsilon_0 Q_0=e/2$. The peak value has a non-monotonic behavior
with $t_\mathrm{FE}$. The inset in panel (b) shows the conductance peak
value, $\sigma_\mathrm{max}=\sigma|_{\epsilon_0Q_0=e/2}$, as a function
of $t_\mathrm{FE}$. The curve has a peak at $t_\mathrm{FE}\approx t_\mathrm d$.
\begin{figure}
\includegraphics[width=1\columnwidth]{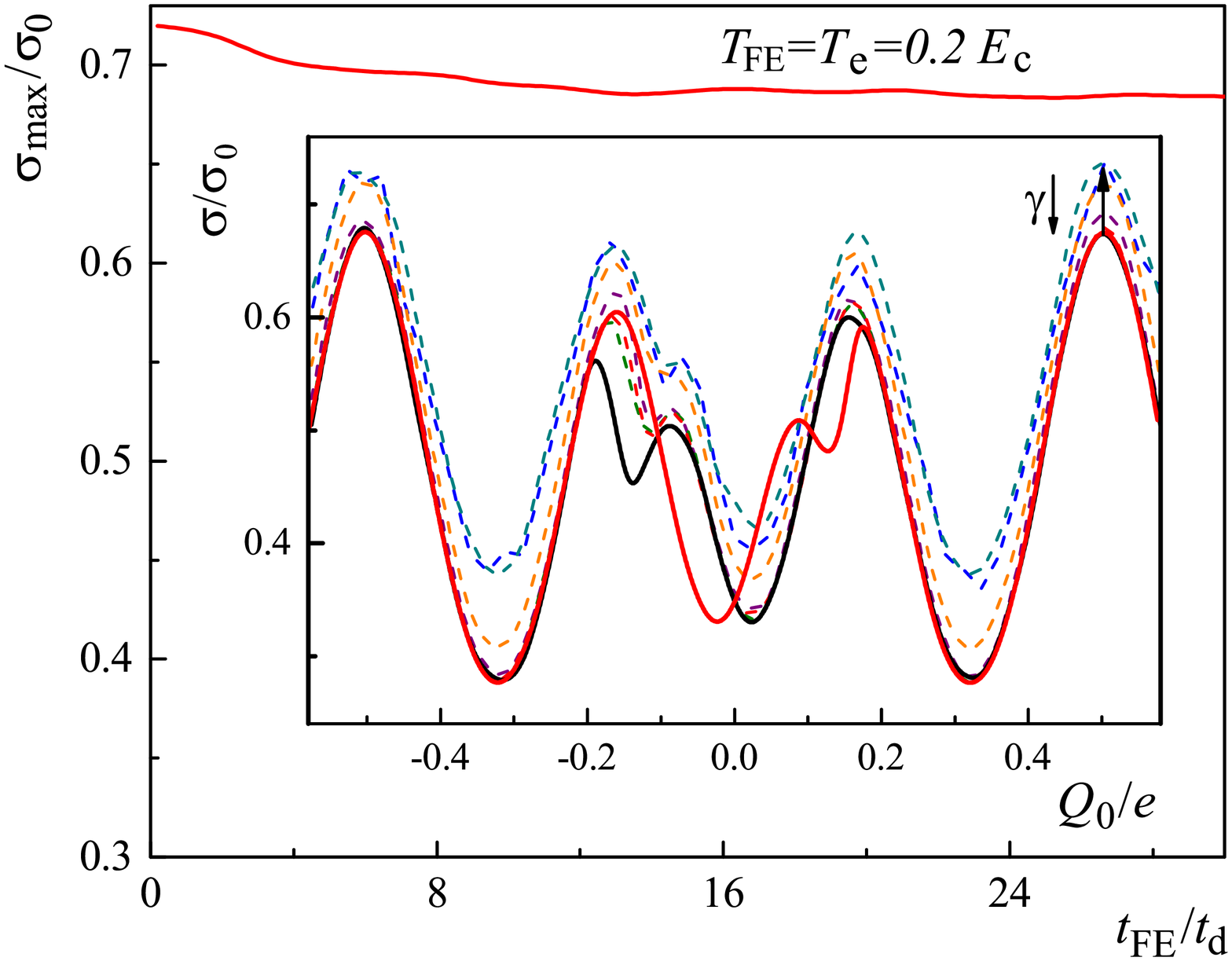}
\caption{(Color online) Conductance, $\sigma_{\mathrm{max}}=\sigma|_{Q_0=e/2}$, vs.
 $t_\mathrm{FE}$ for negative susceptibility, $\chi_0$,
 ($T_\mathrm e=T_\mathrm{FE}<T_\mathrm C^0$). Inset: conductance of SET
 as a function of $Q_0$ for different $t_\mathrm{FE}$. Red solid curve
 stands for increasing $Q_0$. All other curves stand for decreasing $Q_0$.
 Solid red and black lines are for $t_\mathrm{FE}\gg t_\mathrm{d}$. Blue dashed
 line shows the opposite case, $t_\mathrm{FE}/t_\mathrm d=0.1$. All
 other curves correspond to intermediate $t_\mathrm{FE}$.}  \label{Fig:Cond_Qg_gamma_1}
\end{figure}

If susceptibility, $\chi_0$, is small and the hysteresis is absent at $t_\mathrm{FE}\gg t_\mathrm d$,
the conductance is a periodic function of $Q_0$ (in the absence of non-linear
term in Eq.~(\ref{Eq:ShiftFE_NL})). Figure~\ref{Fig:SmallSusc} shows
the dependence of the peak value of the conductance, $\sigma_\mathrm{max}=\sigma|_{\epsilon_0Q_0=e/2}$,
on the FE relaxation time, $t_\mathrm{FE}$. The curves in Fig.~\ref{Fig:SmallSusc} are
plotted in the presence of FE fluctuations for the following parameters:
$\alpha_{\mathrm P}=2$, $\Cg0=0.01 \Cs$, $T_\mathrm e=T_\mathrm{FE}=0.2 E_\mathrm c$.
One can see that in this case the peak value
does not depend on the FE relaxation time, $t_\mathrm{FE}$. The inset
in Fig.~\ref{Fig:SmallSusc} shows the dependence of the conductance peak value
on temperature for time $t_\mathrm{FE}\gg t_\mathrm d$, see the black dashed line.
The peak value increases with temperature. It is in a qualitative agreement with
Eq.~(\ref{Eq:CondPeakAv}). The red line in the inset shows the same function
as in Eq.~(\ref{Eq:CondPeakAv}), however with different coefficient ($1/7$) instead
of ($1/18$). All curves in the inset are plotted for fixed susceptibility, $\chi_0$.

\subsubsection{$T_\mathrm{FE}<T^0_\mathrm C$.}

Above we have shown that for ``fast'' FE the conductance has hysteresis as a function of $Q_0$.
Our MC simulations show that this hysteresis exists even for ``slow'' FE.
The hysteresis is robust against the FE fluctuations (see red and black lines in the inset in Fig.~\ref{Fig:Cond_Qg_gamma_1}).
Inset in Fig.~\ref{Fig:Cond_Qg_gamma_1} shows
the SET conductance versus the gate charge $Q_0$ and
the FE relaxation time $t_\mathrm{FE}$ in the presence of FE polarization fluctuations
for the following parameters: $\Cg0=0.2 \Cs$
and $T_\mathrm e=T_\mathrm{FE}=0.2 E_\mathrm c$, $\alpha_\mathrm P=-2$, $e^2\tilde\beta_\mathrm P=0.13$.
The conductance peak value decreases with increasing the FE relaxation time (see the curve in Fig.~\ref{Fig:Cond_Qg_gamma_1}).

\subsection{Discussion of some experimental problems}

The important question about implementation of the SET with FE layer in the gate
capacitor is related to the FE Curie temperature. This temperature and the
dielectric permittivity decrease with decreasing
the FE layer thickness~[\onlinecite{Frid2006rev},\onlinecite{Frid2010rev}]. A critical thickness exists for most of
FE materials. Below this thickness the FE phase transition does not
exist ($T^0_\mathrm C$ becomes negative). Moreover, boundary conditions at the
FE layer surface in the SET also contribute to the degradation of FE properties.
Since the grain voltage is not fixed the FE internal electric field is not
screened effectively leading to the reduction of the FE temperature and
dielectric permittivity. The question about electric properties of FE material
confined in such a small volume is currently open. In our consideration we
assume FEs with low Curie temperature. One example is TTF-CA FE material.
Bulk TTF-CA under the close circuit boundary conditions has Curie temperature,
$T^0_\mathrm C=80$ K~[\onlinecite{Tokura2012},\onlinecite{Collet2006}]. For this FE material $T_\mathrm C^0$
can be of order of $(0.1-0.2) E_\mathrm c$. In Ref.~[\onlinecite{Plank2014}] TTF-CA film
was confined at one surface with a granular film. Therefore there was no effective
screening of the internal FE field. In this system the FE Curie temperature
has lower value $T^0_\mathrm C=50$ K.

For Coulomb blockade effects to exist at such a high temperature ($\sim 50$ K) it is necessary 
to produce a rather small island in the SET with a size of approximately $5$ nm. 
Assuming that the insulator thickness in the tunnel junctions is about $1$ nm and $\epsilon_{\mathrm{jun}}\approx 5$ 
we find, $C\sim 10^{-18}$ F. Usually, the resistance of the tunnelling junction is 
about $10^6$ Ohm leading to the discharging time $t_\mathrm d\sim10^{-12}$ s. Using tunnel junctions with resistance $\sim 10^7$ Ohm one can increase the FE discharging time to $t_\mathrm d\sim10^{-11}$ s.

It is known that dielectric relaxation time in materials depends on the mechanism of dielectric response. Two groups of materials are most promising for our purposes. In the first group the dielectric response is caused by the elastic mechanisms. The electronic mechanism does not produce a very high dielectric permittivity, instead it gives a very fast response with $t_\mathrm{FE}\sim 10^{-15}$ s. Usual insulators such as Si,  SiO$_2$ and Al$_2$O$_3$ have such a response mechanism. These materials can not be used as ``retarded'' in a SET. A SET with these insulators can 
be described using ``classical'' theory. The ionic  elastic mechanism gives a 
higher dielectric permittivity but slower response with $t_\mathrm{FE}\sim 10^{-11}-10^{-14}$ s. 
This mechanism substantially contributes  to the dielectric response of 
semiconductors (GaAs and CdS) and ionic crystals such as NaCl. Insulators with this type of dielectric response 
can be used to investigate the limits $t_{\mathrm{FE}}\ll t_\mathrm d$ and $t_{\mathrm{FE}}\sim t_\mathrm d$. 
Second group of materials have thermal mechanisms of dielectric response. These 
mechanisms give a rather high permittivity but longer relaxation time, 
$t_\mathrm{FE}\sim 10^{-4}-10^{-10}$ s. Moreover, relaxation time in these materials 
increases with decreasing temperature. The dielectric response of 
relaxor FEs (such as PMN-PSN~[\onlinecite{Sternberg2010}] with $t_\mathrm{FE}\sim 10^{-9}$ s at room temperature) 
are usually due to thermal mechanism. These materials can be considered as slow FEs with $t_{\mathrm{FE}}\gg t_\mathrm d$.

Most FEs materials are rather slow with dielectric relaxation time well below 1 THz. This is due to ionic nature of their dielectric response. However, recently new types of electronic FEs were discovered. One of them is TTF-CA. This material has dielectric relaxation time in the THz region~[\onlinecite{Okamoto2013}]. This material can be used 
to study interaction of SET with retarded FEs.

\section{Conclusion}

We studied the SET with a FE or a
dielectric placed in the gate capacitor. In  contrast to the previous papers,
we investigated the system behavior for arbitrary ratio between the FE
relaxation time, $t_\mathrm{FE}$, and the SET characteristic time, $t_\mathrm d$.
We used analytical methods and Monte-Carlo simulations to study the SET behavior.

We considered the case of ``fast'' FE (dielectric) with $\hbar /E_\mathrm c\ll t_\mathrm{FE} \ll t_\mathrm d$.
We showed that this case is different from the ``classical'' SET theory, where
the instant response, $\hbar /E_\mathrm c\gg t_\mathrm{FE}$, of dielectric material is
assumed. Thus, ``fast'' FE has the retarded response.
We showed that retarded FE (dielectric) response leads to the appearance of
an additional contribution to the Coulomb blockade effect and suppresses the
SET conductance. The conductance as a function of the gate charge $Q_0$ behaves
differently at different temperatures. Below a certain temperature $T_\mathrm C^0$
(which is the Curie temperature of the FE inside the gate capacitor)
the conductance shows the hysteresis behavior.
Above $T_\mathrm C^0$ no hysteresis appears in the case of ``fast'' FE.
Monte-Carlo simulations allowed us to take into account fluctuations of the FE (dielectric) polarization.
For ``fast'' FE these fluctuations partially suppress the additional Coulomb blockade effect.
We showed that conductance hysteresis is robust against these fluctuations.

We studied the case of ``slow'' FE (dielectric), $t_\mathrm{FE} \gg t_\mathrm d$.
This limit was discussed in literature using the mean field approach in the past.
The interesting feature is related to the fact that due to coupling
of fast SET subsystem with slow FE subsystem the conductance of the SET showed the
hysteresis behavior even for temperatures, $T>T_\mathrm C^0$. Thus,
the nature of hysteresis is not related to the spontaneous polarization
of the FE layer and may appear even with simple dielectric being placed in the gate capacitor.
We used Monte-Carlo simulations to re-investigate this case. We found that
FE (dielectric) fluctuations suppress the hysteresis. The only way to observe this hysteresis
is to produce the FE (dielectric) layer temperature lower than the temperature of electrons
in the SET leads (source and drain) and the island, ($T_\mathrm{FE}\ll T_\mathrm e$).

We showed that for ``slow'' FE and low temperature, $T<T^0_\mathrm C$, the SET conductance
as function of the gate charge $Q_0$ shows the hysteresis due to FE instability.
This hysteresis is robust against FE fluctuations.

Using the Monte-Carlo simulations we studied the transitional region between
``fast'' and ``slow'' FE. The transition depends on temperature (or the susceptibility of the
FE (dielectric) layer). At high temperatures (low susceptibility) the peak value of the SET
conductance is almost independent of the FE relaxation time.
For temperatures close to $T^0_\mathrm C$ (high susceptibility) the conductance peak value
non-monotonically depends on the FE relaxation time. A maximum appears
close to $t_\mathrm{FE}=t_\mathrm d$. Below the Curie point ($T<T^0_\mathrm C$)
the conductance peak value decreases with increasing $t_\mathrm{FE}$.

\section{Acknowledgements}

I.~B. was supported by NSF under Cooperative Agreement Award EEC-1160504, NSF PREM Award
and the U.S. Civilian Research and Development Foundation (CRDF Global). N. C. and S. F. are grateful to Russian Academy of Sciences for the access to JSCC and �Uran� clusters and Kurchatov center for access to HCP supercomputer cluster. S. F. acknowledges support of the Russian foundation for Basic Research (grant No. 13-02-00579). N. C. acknowledges  Russian Science Foundation (grant No. 14-12-01185) for support of supercomputer simulations.

\bibliography{TDSET}

%merlin.mbs apsrev4-1.bst 2010-07-25 4.21a (PWD, AO, DPC) hacked
%Control: key (0)
%Control: author (72) initials jnrlst
%Control: editor formatted (1) identically to author
%Control: production of article title (-1) disabled
%Control: page (0) single
%Control: year (1) truncated
%Control: production of eprint (0) enabled
\begin{thebibliography}{52}%
\makeatletter
\providecommand \@ifxundefined [1]{%
 \@ifx{#1\undefined}
}%
\providecommand \@ifnum [1]{%
 \ifnum #1\expandafter \@firstoftwo
 \else \expandafter \@secondoftwo
 \fi
}%
\providecommand \@ifx [1]{%
 \ifx #1\expandafter \@firstoftwo
 \else \expandafter \@secondoftwo
 \fi
}%
\providecommand \natexlab [1]{#1}%
\providecommand \enquote  [1]{``#1''}%
\providecommand \bibnamefont  [1]{#1}%
\providecommand \bibfnamefont [1]{#1}%
\providecommand \citenamefont [1]{#1}%
\providecommand \href@noop [0]{\@secondoftwo}%
\providecommand \href [0]{\begingroup \@sanitize@url \@href}%
\providecommand \@href[1]{\@@startlink{#1}\@@href}%
\providecommand \@@href[1]{\endgroup#1\@@endlink}%
\providecommand \@sanitize@url [0]{\catcode `\\12\catcode `\$12\catcode
  `\&12\catcode `\#12\catcode `\^12\catcode `\_12\catcode `\%12\relax}%
\providecommand \@@startlink[1]{}%
\providecommand \@@endlink[0]{}%
\providecommand \url  [0]{\begingroup\@sanitize@url \@url }%
\providecommand \@url [1]{\endgroup\@href {#1}{\urlprefix }}%
\providecommand \urlprefix  [0]{URL }%
\providecommand \Eprint [0]{\href }%
\providecommand \doibase [0]{http://dx.doi.org/}%
\providecommand \selectlanguage [0]{\@gobble}%
\providecommand \bibinfo  [0]{\@secondoftwo}%
\providecommand \bibfield  [0]{\@secondoftwo}%
\providecommand \translation [1]{[#1]}%
\providecommand \BibitemOpen [0]{}%
\providecommand \bibitemStop [0]{}%
\providecommand \bibitemNoStop [0]{.\EOS\space}%
\providecommand \EOS [0]{\spacefactor3000\relax}%
\providecommand \BibitemShut  [1]{\csname bibitem#1\endcsname}%
\let\auto@bib@innerbib\@empty
%</preamble>
\bibitem [{\citenamefont {Giazotto}(2015)}]{Giazotto2015}%
  \BibitemOpen
  \bibfield  {author} {\bibinfo {author} {\bibfnamefont {F.}~\bibnamefont
  {Giazotto}},\ }\href@noop {} {\bibfield  {journal} {\bibinfo  {journal}
  {Nature Phys.}\ }\textbf {\bibinfo {volume} {11}},\ \bibinfo {pages} {527}
  (\bibinfo {year} {2015})}\BibitemShut {NoStop}%
\bibitem [{\citenamefont {van Woerkom}\ \emph {et~al.}(2015)\citenamefont {van
  Woerkom}, \citenamefont {Geresdi},\ and\ \citenamefont
  {Kouwenhoven}}]{Kouwenhoven2015}%
  \BibitemOpen
  \bibfield  {author} {\bibinfo {author} {\bibfnamefont {D.~J.}\ \bibnamefont
  {van Woerkom}}, \bibinfo {author} {\bibfnamefont {A.}~\bibnamefont
  {Geresdi}}, \ and\ \bibinfo {author} {\bibfnamefont {L.~P.}\ \bibnamefont
  {Kouwenhoven}},\ }\href@noop {} {\bibfield  {journal} {\bibinfo  {journal}
  {Nature Phys.}\ }\textbf {\bibinfo {volume} {11}},\ \bibinfo {pages} {547}
  (\bibinfo {year} {2015})}\BibitemShut {NoStop}%
\bibitem [{\citenamefont {Li}\ \emph {et~al.}(2015)\citenamefont {Li},
  \citenamefont {Lam},\ and\ \citenamefont {You}}]{You2015}%
  \BibitemOpen
  \bibfield  {author} {\bibinfo {author} {\bibfnamefont {Z.-Z.}\ \bibnamefont
  {Li}}, \bibinfo {author} {\bibfnamefont {C.-H.}\ \bibnamefont {Lam}}, \ and\
  \bibinfo {author} {\bibfnamefont {J.~Q.}\ \bibnamefont {You}},\ }\href@noop
  {} {\bibfield  {journal} {\bibinfo  {journal} {Scientific Reports}\ }\textbf
  {\bibinfo {volume} {5}},\ \bibinfo {pages} {11416} (\bibinfo {year}
  {2015})}\BibitemShut {NoStop}%
\bibitem [{\citenamefont {Khaimovich}\ \emph {et~al.}(2015)\citenamefont
  {Khaimovich}, \citenamefont {Koski}, \citenamefont {Saira}, \citenamefont
  {Kravtsov},\ and\ \citenamefont {Pekola}}]{Pekola2015}%
  \BibitemOpen
  \bibfield  {author} {\bibinfo {author} {\bibfnamefont {I.~M.}\ \bibnamefont
  {Khaimovich}}, \bibinfo {author} {\bibfnamefont {J.~V.}\ \bibnamefont
  {Koski}}, \bibinfo {author} {\bibfnamefont {O.~P.}\ \bibnamefont {Saira}},
  \bibinfo {author} {\bibfnamefont {V.~E.}\ \bibnamefont {Kravtsov}}, \ and\
  \bibinfo {author} {\bibfnamefont {J.~P.}\ \bibnamefont {Pekola}},\
  }\href@noop {} {\bibfield  {journal} {\bibinfo  {journal} {Nature
  Communications}\ }\textbf {\bibinfo {volume} {6}},\ \bibinfo {pages} {7010}
  (\bibinfo {year} {2015})}\BibitemShut {NoStop}%
\bibitem [{\citenamefont {Horibe}\ \emph {et~al.}(2015)\citenamefont {Horibe},
  \citenamefont {Kodera},\ and\ \citenamefont {Oda}}]{Oda2015}%
  \BibitemOpen
  \bibfield  {author} {\bibinfo {author} {\bibfnamefont {K.}~\bibnamefont
  {Horibe}}, \bibinfo {author} {\bibfnamefont {T.}~\bibnamefont {Kodera}}, \
  and\ \bibinfo {author} {\bibfnamefont {S.}~\bibnamefont {Oda}},\ }\href@noop
  {} {\bibfield  {journal} {\bibinfo  {journal} {Appl. Phys. Lett.}\ }\textbf
  {\bibinfo {volume} {106}},\ \bibinfo {pages} {053119} (\bibinfo {year}
  {2015})}\BibitemShut {NoStop}%
\bibitem [{\citenamefont {Islam}\ \emph {et~al.}(2015)\citenamefont {Islam},
  \citenamefont {Joung},\ and\ \citenamefont {Khondaker}}]{Khondaker2015}%
  \BibitemOpen
  \bibfield  {author} {\bibinfo {author} {\bibfnamefont {M.~R.}\ \bibnamefont
  {Islam}}, \bibinfo {author} {\bibfnamefont {D.}~\bibnamefont {Joung}}, \ and\
  \bibinfo {author} {\bibfnamefont {S.~I.}\ \bibnamefont {Khondaker}},\
  }\href@noop {} {\bibfield  {journal} {\bibinfo  {journal} {Nanoscale}\
  }\textbf {\bibinfo {volume} {7}},\ \bibinfo {pages} {9786} (\bibinfo {year}
  {2015})}\BibitemShut {NoStop}%
\bibitem [{\citenamefont {Hakkinen}\ \emph {et~al.}(2015)\citenamefont
  {Hakkinen}, \citenamefont {Isacsson}, \citenamefont {Savin}, \citenamefont
  {Sulkko},\ and\ \citenamefont {Hakkonen}}]{Hakonen2015}%
  \BibitemOpen
  \bibfield  {author} {\bibinfo {author} {\bibfnamefont {P.}~\bibnamefont
  {Hakkinen}}, \bibinfo {author} {\bibfnamefont {A.}~\bibnamefont {Isacsson}},
  \bibinfo {author} {\bibfnamefont {A.}~\bibnamefont {Savin}}, \bibinfo
  {author} {\bibfnamefont {J.}~\bibnamefont {Sulkko}}, \ and\ \bibinfo {author}
  {\bibfnamefont {P.}~\bibnamefont {Hakkonen}},\ }\href@noop {} {\bibfield
  {journal} {\bibinfo  {journal} {Nano Lett.}\ }\textbf {\bibinfo {volume}
  {15}},\ \bibinfo {pages} {1667} (\bibinfo {year} {2015})}\BibitemShut
  {NoStop}%
\bibitem [{\citenamefont {Strasberg}\ \emph {et~al.}(2013)\citenamefont
  {Strasberg}, \citenamefont {Schaller}, \citenamefont {Brandes},\ and\
  \citenamefont {Esposito}}]{Esposito2013}%
  \BibitemOpen
  \bibfield  {author} {\bibinfo {author} {\bibfnamefont {P.}~\bibnamefont
  {Strasberg}}, \bibinfo {author} {\bibfnamefont {G.}~\bibnamefont {Schaller}},
  \bibinfo {author} {\bibfnamefont {T.}~\bibnamefont {Brandes}}, \ and\
  \bibinfo {author} {\bibfnamefont {M.}~\bibnamefont {Esposito}},\ }\href@noop
  {} {\bibfield  {journal} {\bibinfo  {journal} {Phys. Rev. Lett.}\ }\textbf
  {\bibinfo {volume} {110}},\ \bibinfo {pages} {040601} (\bibinfo {year}
  {2013})}\BibitemShut {NoStop}%
\bibitem [{\citenamefont {Feshchenko}\ \emph {et~al.}(2014)\citenamefont
  {Feshchenko}, \citenamefont {Koski},\ and\ \citenamefont
  {Pekola}}]{Pekola2014}%
  \BibitemOpen
  \bibfield  {author} {\bibinfo {author} {\bibfnamefont {A.~V.}\ \bibnamefont
  {Feshchenko}}, \bibinfo {author} {\bibfnamefont {J.~V.}\ \bibnamefont
  {Koski}}, \ and\ \bibinfo {author} {\bibfnamefont {J.~P.}\ \bibnamefont
  {Pekola}},\ }\href@noop {} {\bibfield  {journal} {\bibinfo  {journal} {Phys.
  Rev. B}\ }\textbf {\bibinfo {volume} {90}},\ \bibinfo {pages} {201407(R)}
  (\bibinfo {year} {2014})}\BibitemShut {NoStop}%
\bibitem [{\citenamefont {Kirton}\ and\ \citenamefont
  {Armour}(2013)}]{Armour2013}%
  \BibitemOpen
  \bibfield  {author} {\bibinfo {author} {\bibfnamefont {P.~G.}\ \bibnamefont
  {Kirton}}\ and\ \bibinfo {author} {\bibfnamefont {A.~D.}\ \bibnamefont
  {Armour}},\ }\href@noop {} {\bibfield  {journal} {\bibinfo  {journal} {Phys.
  Rev. B}\ }\textbf {\bibinfo {volume} {87}},\ \bibinfo {pages} {155407}
  (\bibinfo {year} {2013})}\BibitemShut {NoStop}%
\bibitem [{\citenamefont {Likharev}\ \emph {et~al.}(1989)\citenamefont
  {Likharev}, \citenamefont {Bakhvalov}, \citenamefont {Kazacha},\ and\
  \citenamefont {Serdyukova}}]{Serdyukova1989}%
  \BibitemOpen
  \bibfield  {author} {\bibinfo {author} {\bibfnamefont {K.~K.}\ \bibnamefont
  {Likharev}}, \bibinfo {author} {\bibfnamefont {N.~S.}\ \bibnamefont
  {Bakhvalov}}, \bibinfo {author} {\bibfnamefont {G.~S.}\ \bibnamefont
  {Kazacha}}, \ and\ \bibinfo {author} {\bibfnamefont {S.~I.}\ \bibnamefont
  {Serdyukova}},\ }\href@noop {} {\bibfield  {journal} {\bibinfo  {journal}
  {IEEE Trans. Magn.}\ }\textbf {\bibinfo {volume} {25}},\ \bibinfo {pages}
  {1436} (\bibinfo {year} {1989})}\BibitemShut {NoStop}%
\bibitem [{\citenamefont {Averin}\ and\ \citenamefont
  {Likharev}(1986)}]{Likharev1986}%
  \BibitemOpen
  \bibfield  {author} {\bibinfo {author} {\bibfnamefont {D.~V.}\ \bibnamefont
  {Averin}}\ and\ \bibinfo {author} {\bibfnamefont {K.~K.}\ \bibnamefont
  {Likharev}},\ }\href@noop {} {\bibfield  {journal} {\bibinfo  {journal}
  {Journal of Low Temperature Physics}\ }\textbf {\bibinfo {volume} {62}},\
  \bibinfo {pages} {345} (\bibinfo {year} {1986})}\BibitemShut {NoStop}%
\bibitem [{\citenamefont {Gorelik}\ \emph {et~al.}(1998)\citenamefont
  {Gorelik}, \citenamefont {Isacsson}, \citenamefont {Voinova}, \citenamefont
  {Kasemo}, \citenamefont {Shekhter},\ and\ \citenamefont
  {Jonson}}]{Jonson1998}%
  \BibitemOpen
  \bibfield  {author} {\bibinfo {author} {\bibfnamefont {L.~Y.}\ \bibnamefont
  {Gorelik}}, \bibinfo {author} {\bibfnamefont {A.}~\bibnamefont {Isacsson}},
  \bibinfo {author} {\bibfnamefont {M.~V.}\ \bibnamefont {Voinova}}, \bibinfo
  {author} {\bibfnamefont {B.}~\bibnamefont {Kasemo}}, \bibinfo {author}
  {\bibfnamefont {R.~I.}\ \bibnamefont {Shekhter}}, \ and\ \bibinfo {author}
  {\bibfnamefont {M.}~\bibnamefont {Jonson}},\ }\href@noop {} {\bibfield
  {journal} {\bibinfo  {journal} {Phys. Rev. Lett.}\ }\textbf {\bibinfo
  {volume} {80}},\ \bibinfo {pages} {4526} (\bibinfo {year}
  {1998})}\BibitemShut {NoStop}%
\bibitem [{\citenamefont {Armour}\ and\ \citenamefont
  {MacKinnon}(2002)}]{MacKinnon2002}%
  \BibitemOpen
  \bibfield  {author} {\bibinfo {author} {\bibfnamefont {A.~D.}\ \bibnamefont
  {Armour}}\ and\ \bibinfo {author} {\bibfnamefont {A.}~\bibnamefont
  {MacKinnon}},\ }\href@noop {} {\bibfield  {journal} {\bibinfo  {journal}
  {Phys. Rev. B}\ }\textbf {\bibinfo {volume} {66}},\ \bibinfo {pages} {035333}
  (\bibinfo {year} {2002})}\BibitemShut {NoStop}%
\bibitem [{\citenamefont {Nord}\ \emph {et~al.}(2002)\citenamefont {Nord},
  \citenamefont {Gorelik}, \citenamefont {Shekhter},\ and\ \citenamefont
  {Jonson}}]{Jonson2002}%
  \BibitemOpen
  \bibfield  {author} {\bibinfo {author} {\bibfnamefont {T.}~\bibnamefont
  {Nord}}, \bibinfo {author} {\bibfnamefont {L.~Y.}\ \bibnamefont {Gorelik}},
  \bibinfo {author} {\bibfnamefont {R.~I.}\ \bibnamefont {Shekhter}}, \ and\
  \bibinfo {author} {\bibfnamefont {M.}~\bibnamefont {Jonson}},\ }\href@noop {}
  {\bibfield  {journal} {\bibinfo  {journal} {Phys. Rev. B}\ }\textbf {\bibinfo
  {volume} {65}},\ \bibinfo {pages} {165312} (\bibinfo {year}
  {2002})}\BibitemShut {NoStop}%
\bibitem [{\citenamefont {Novotny}\ \emph {et~al.}(2003)\citenamefont
  {Novotny}, \citenamefont {Donarini},\ and\ \citenamefont
  {Jauho}}]{Jauho2003}%
  \BibitemOpen
  \bibfield  {author} {\bibinfo {author} {\bibfnamefont {T.}~\bibnamefont
  {Novotny}}, \bibinfo {author} {\bibfnamefont {A.}~\bibnamefont {Donarini}}, \
  and\ \bibinfo {author} {\bibfnamefont {A.-P.}\ \bibnamefont {Jauho}},\
  }\href@noop {} {\bibfield  {journal} {\bibinfo  {journal} {Phys. Rev. Lett.}\
  }\textbf {\bibinfo {volume} {90}},\ \bibinfo {pages} {256801} (\bibinfo
  {year} {2003})}\BibitemShut {NoStop}%
\bibitem [{\citenamefont {Chtchelkatchev}\ \emph {et~al.}(2004)\citenamefont
  {Chtchelkatchev}, \citenamefont {Belzig},\ and\ \citenamefont
  {Bruder}}]{Bruder2004}%
  \BibitemOpen
  \bibfield  {author} {\bibinfo {author} {\bibfnamefont {N.~M.}\ \bibnamefont
  {Chtchelkatchev}}, \bibinfo {author} {\bibfnamefont {W.}~\bibnamefont
  {Belzig}}, \ and\ \bibinfo {author} {\bibfnamefont {C.}~\bibnamefont
  {Bruder}},\ }\href@noop {} {\bibfield  {journal} {\bibinfo  {journal} {Phys.
  Rev. B}\ }\textbf {\bibinfo {volume} {70}},\ \bibinfo {pages} {193305}
  (\bibinfo {year} {2004})}\BibitemShut {NoStop}%
\bibitem [{\citenamefont {Armour}\ \emph {et~al.}(2004)\citenamefont {Armour},
  \citenamefont {Blencowe},\ and\ \citenamefont {Zhang}}]{Zhang2004}%
  \BibitemOpen
  \bibfield  {author} {\bibinfo {author} {\bibfnamefont {A.~D.}\ \bibnamefont
  {Armour}}, \bibinfo {author} {\bibfnamefont {M.~P.}\ \bibnamefont
  {Blencowe}}, \ and\ \bibinfo {author} {\bibfnamefont {Y.}~\bibnamefont
  {Zhang}},\ }\href@noop {} {\bibfield  {journal} {\bibinfo  {journal} {Phys.
  Rev. B}\ }\textbf {\bibinfo {volume} {69}},\ \bibinfo {pages} {125313}
  (\bibinfo {year} {2004})}\BibitemShut {NoStop}%
\bibitem [{\citenamefont {Armour}(2004)}]{Armour2004}%
  \BibitemOpen
  \bibfield  {author} {\bibinfo {author} {\bibfnamefont {A.~D.}\ \bibnamefont
  {Armour}},\ }\href@noop {} {\bibfield  {journal} {\bibinfo  {journal} {Phys.
  Rev. B}\ }\textbf {\bibinfo {volume} {70}},\ \bibinfo {pages} {165315}
  (\bibinfo {year} {2004})}\BibitemShut {NoStop}%
\bibitem [{\citenamefont {Rodrigues}\ \emph {et~al.}(2007)\citenamefont
  {Rodrigues}, \citenamefont {Imbers},\ and\ \citenamefont
  {Armour}}]{Armour2007}%
  \BibitemOpen
  \bibfield  {author} {\bibinfo {author} {\bibfnamefont {D.~A.}\ \bibnamefont
  {Rodrigues}}, \bibinfo {author} {\bibfnamefont {J.}~\bibnamefont {Imbers}}, \
  and\ \bibinfo {author} {\bibfnamefont {A.~D.}\ \bibnamefont {Armour}},\
  }\href@noop {} {\bibfield  {journal} {\bibinfo  {journal} {Phys. Rev. Lett.}\
  }\textbf {\bibinfo {volume} {98}},\ \bibinfo {pages} {067204} (\bibinfo
  {year} {2007})}\BibitemShut {NoStop}%
\bibitem [{\citenamefont {Doiron}\ \emph {et~al.}(2006)\citenamefont {Doiron},
  \citenamefont {Belzig},\ and\ \citenamefont {Bruder}}]{Bruder2006}%
  \BibitemOpen
  \bibfield  {author} {\bibinfo {author} {\bibfnamefont {C.~B.}\ \bibnamefont
  {Doiron}}, \bibinfo {author} {\bibfnamefont {W.}~\bibnamefont {Belzig}}, \
  and\ \bibinfo {author} {\bibfnamefont {C.}~\bibnamefont {Bruder}},\
  }\href@noop {} {\bibfield  {journal} {\bibinfo  {journal} {Phys. Rev. B}\
  }\textbf {\bibinfo {volume} {74}},\ \bibinfo {pages} {205336} (\bibinfo
  {year} {2006})}\BibitemShut {NoStop}%
\bibitem [{\citenamefont {Hussein}\ \emph {et~al.}(2010)\citenamefont
  {Hussein}, \citenamefont {Metelmann}, \citenamefont {Zedler},\ and\
  \citenamefont {Brandes}}]{Brandes2010}%
  \BibitemOpen
  \bibfield  {author} {\bibinfo {author} {\bibfnamefont {R.}~\bibnamefont
  {Hussein}}, \bibinfo {author} {\bibfnamefont {A.}~\bibnamefont {Metelmann}},
  \bibinfo {author} {\bibfnamefont {P.}~\bibnamefont {Zedler}}, \ and\ \bibinfo
  {author} {\bibfnamefont {T.}~\bibnamefont {Brandes}},\ }\href@noop {}
  {\bibfield  {journal} {\bibinfo  {journal} {Phys. Rev. B}\ }\textbf {\bibinfo
  {volume} {82}},\ \bibinfo {pages} {165406} (\bibinfo {year}
  {2010})}\BibitemShut {NoStop}%
\bibitem [{\citenamefont {Fedorov}\ \emph
  {et~al.}(2014{\natexlab{a}})\citenamefont {Fedorov}, \citenamefont
  {Korolkov}, \citenamefont {Chtchelkatchev}, \citenamefont {Udalov},\ and\
  \citenamefont {Beloborodov}}]{Beloborodov2014}%
  \BibitemOpen
  \bibfield  {author} {\bibinfo {author} {\bibfnamefont {S.~A.}\ \bibnamefont
  {Fedorov}}, \bibinfo {author} {\bibfnamefont {A.~E.}\ \bibnamefont
  {Korolkov}}, \bibinfo {author} {\bibfnamefont {N.~M.}\ \bibnamefont
  {Chtchelkatchev}}, \bibinfo {author} {\bibfnamefont {O.~G.}\ \bibnamefont
  {Udalov}}, \ and\ \bibinfo {author} {\bibfnamefont {I.~S.}\ \bibnamefont
  {Beloborodov}},\ }\href {\doibase 10.1103/PhysRevB.90.195111} {\bibfield
  {journal} {\bibinfo  {journal} {Phys. Rev. B}\ }\textbf {\bibinfo {volume}
  {90}},\ \bibinfo {pages} {195111} (\bibinfo {year}
  {2014}{\natexlab{a}})}\BibitemShut {NoStop}%
\bibitem [{\citenamefont {Fedorov}\ \emph
  {et~al.}(2014{\natexlab{b}})\citenamefont {Fedorov}, \citenamefont
  {Korolkov}, \citenamefont {Chtchelkatchev}, \citenamefont {Udalov},\ and\
  \citenamefont {Beloborodov}}]{Beloborodov2014_1}%
  \BibitemOpen
  \bibfield  {author} {\bibinfo {author} {\bibfnamefont {S.~A.}\ \bibnamefont
  {Fedorov}}, \bibinfo {author} {\bibfnamefont {A.~E.}\ \bibnamefont
  {Korolkov}}, \bibinfo {author} {\bibfnamefont {N.~M.}\ \bibnamefont
  {Chtchelkatchev}}, \bibinfo {author} {\bibfnamefont {O.~G.}\ \bibnamefont
  {Udalov}}, \ and\ \bibinfo {author} {\bibfnamefont {I.~S.}\ \bibnamefont
  {Beloborodov}},\ }\href {\doibase 10.1103/PhysRevB.89.155410} {\bibfield
  {journal} {\bibinfo  {journal} {Phys. Rev. B}\ }\textbf {\bibinfo {volume}
  {89}},\ \bibinfo {pages} {155410} (\bibinfo {year}
  {2014}{\natexlab{b}})}\BibitemShut {NoStop}%
\bibitem [{\citenamefont {Pekola}\ \emph {et~al.}(2014)\citenamefont {Pekola},
  \citenamefont {Vartiainen}, \citenamefont {Mottonen}, \citenamefont {Saira},
  \citenamefont {Meschke},\ and\ \citenamefont {Averin}}]{Averin2014}%
  \BibitemOpen
  \bibfield  {author} {\bibinfo {author} {\bibfnamefont {J.~P.}\ \bibnamefont
  {Pekola}}, \bibinfo {author} {\bibfnamefont {J.~J.}\ \bibnamefont
  {Vartiainen}}, \bibinfo {author} {\bibfnamefont {M.}~\bibnamefont
  {Mottonen}}, \bibinfo {author} {\bibfnamefont {O.-P.}\ \bibnamefont {Saira}},
  \bibinfo {author} {\bibfnamefont {M.}~\bibnamefont {Meschke}}, \ and\
  \bibinfo {author} {\bibfnamefont {D.~V.}\ \bibnamefont {Averin}},\
  }\href@noop {} {\bibfield  {journal} {\bibinfo  {journal} {Nature Physics}\
  }\textbf {\bibinfo {volume} {4}},\ \bibinfo {pages} {120} (\bibinfo {year}
  {2014})}\BibitemShut {NoStop}%
\bibitem [{\citenamefont {Strukov}\ and\ \citenamefont
  {Levanyuk}(1998)}]{Levan1983}%
  \BibitemOpen
  \bibfield  {author} {\bibinfo {author} {\bibfnamefont {B.~A.}\ \bibnamefont
  {Strukov}}\ and\ \bibinfo {author} {\bibfnamefont {A.~P.}\ \bibnamefont
  {Levanyuk}},\ }\href@noop {} {\emph {\bibinfo {title} {Ferroelectric
  Phenomena in Crystals}}}\ (\bibinfo  {publisher} {Springer, Geidelberg,
  1998},\ \bibinfo {year} {1998})\BibitemShut {NoStop}%
\bibitem [{\citenamefont {Fedorov}\ \emph {et~al.}(2015)\citenamefont
  {Fedorov}, \citenamefont {Chtchelkatchev}, \citenamefont {Udalov},\ and\
  \citenamefont {Beloborodov}}]{Beloborodov2014_2}%
  \BibitemOpen
  \bibfield  {author} {\bibinfo {author} {\bibfnamefont {S.~A.}\ \bibnamefont
  {Fedorov}}, \bibinfo {author} {\bibfnamefont {N.~M.}\ \bibnamefont
  {Chtchelkatchev}}, \bibinfo {author} {\bibfnamefont {O.~G.}\ \bibnamefont
  {Udalov}}, \ and\ \bibinfo {author} {\bibfnamefont {I.~S.}\ \bibnamefont
  {Beloborodov}},\ }\href@noop {} {\bibfield  {journal} {\bibinfo  {journal}
  {Phys. Rev. B}\ }\textbf {\bibinfo {volume} {92}},\ \bibinfo {pages} {115425}
  (\bibinfo {year} {2015})}\BibitemShut {NoStop}%
\bibitem [{\citenamefont {Fonseca}\ \emph {et~al.}(1995)\citenamefont
  {Fonseca}, \citenamefont {Korotkov}, \citenamefont {Likharev},\ and\
  \citenamefont {Odintsov}}]{Odintsov1995}%
  \BibitemOpen
  \bibfield  {author} {\bibinfo {author} {\bibfnamefont {L.~R.~C.}\
  \bibnamefont {Fonseca}}, \bibinfo {author} {\bibfnamefont {A.~N.}\
  \bibnamefont {Korotkov}}, \bibinfo {author} {\bibfnamefont {K.~K.}\
  \bibnamefont {Likharev}}, \ and\ \bibinfo {author} {\bibfnamefont {A.~A.}\
  \bibnamefont {Odintsov}},\ }\href@noop {} {\bibfield  {journal} {\bibinfo
  {journal} {J. Appl. Phys.}\ }\textbf {\bibinfo {volume} {78}},\ \bibinfo
  {pages} {3238} (\bibinfo {year} {1995})}\BibitemShut {NoStop}%
\bibitem [{\citenamefont {Goan}(2004)}]{Goan2004}%
  \BibitemOpen
  \bibfield  {author} {\bibinfo {author} {\bibfnamefont {H.-S.}\ \bibnamefont
  {Goan}},\ }\href@noop {} {\bibfield  {journal} {\bibinfo  {journal} {Phys.
  Rev. B}\ }\textbf {\bibinfo {volume} {70}},\ \bibinfo {pages} {075305}
  (\bibinfo {year} {2004})}\BibitemShut {NoStop}%
\bibitem [{\citenamefont {Mozyrsky}\ \emph {et~al.}(2004)\citenamefont
  {Mozyrsky}, \citenamefont {Martin},\ and\ \citenamefont
  {Hastings}}]{Hastings2004}%
  \BibitemOpen
  \bibfield  {author} {\bibinfo {author} {\bibfnamefont {D.}~\bibnamefont
  {Mozyrsky}}, \bibinfo {author} {\bibfnamefont {I.}~\bibnamefont {Martin}}, \
  and\ \bibinfo {author} {\bibfnamefont {M.~B.}\ \bibnamefont {Hastings}},\
  }\href@noop {} {\bibfield  {journal} {\bibinfo  {journal} {Phys. Rev. Lett.}\
  }\textbf {\bibinfo {volume} {92}},\ \bibinfo {pages} {018303} (\bibinfo
  {year} {2004})}\BibitemShut {NoStop}%
\bibitem [{\citenamefont {Waahuber}\ and\ \citenamefont
  {Kosina}(1997)}]{Kosina1997}%
  \BibitemOpen
  \bibfield  {author} {\bibinfo {author} {\bibfnamefont {C.}~\bibnamefont
  {Waahuber}}\ and\ \bibinfo {author} {\bibfnamefont {H.}~\bibnamefont
  {Kosina}},\ }\href@noop {} {\bibfield  {journal} {\bibinfo  {journal} {IEEE
  Trans. Computer-Aided Design of Integrated Circuits and Systems}\ }\textbf
  {\bibinfo {volume} {16}},\ \bibinfo {pages} {937} (\bibinfo {year}
  {1997})}\BibitemShut {NoStop}%
\bibitem [{\citenamefont {Bylander}\ \emph {et~al.}(2005)\citenamefont
  {Bylander}, \citenamefont {Duty},\ and\ \citenamefont
  {Delsing}}]{Delsing2005}%
  \BibitemOpen
  \bibfield  {author} {\bibinfo {author} {\bibfnamefont {J.}~\bibnamefont
  {Bylander}}, \bibinfo {author} {\bibfnamefont {T.}~\bibnamefont {Duty}}, \
  and\ \bibinfo {author} {\bibfnamefont {P.}~\bibnamefont {Delsing}},\
  }\href@noop {} {\bibfield  {journal} {\bibinfo  {journal} {Nature}\ }\textbf
  {\bibinfo {volume} {434}},\ \bibinfo {pages} {361} (\bibinfo {year}
  {2005})}\BibitemShut {NoStop}%
\bibitem [{\citenamefont {Giazotto}\ \emph {et~al.}(2006)\citenamefont
  {Giazotto}, \citenamefont {Heikkila}, \citenamefont {Luukanen}, \citenamefont
  {Savin},\ and\ \citenamefont {Pekola}}]{Luukanen2006}%
  \BibitemOpen
  \bibfield  {author} {\bibinfo {author} {\bibfnamefont {F.}~\bibnamefont
  {Giazotto}}, \bibinfo {author} {\bibfnamefont {T.~T.}\ \bibnamefont
  {Heikkila}}, \bibinfo {author} {\bibfnamefont {A.}~\bibnamefont {Luukanen}},
  \bibinfo {author} {\bibfnamefont {A.~M.}\ \bibnamefont {Savin}}, \ and\
  \bibinfo {author} {\bibfnamefont {J.~P.}\ \bibnamefont {Pekola}},\
  }\href@noop {} {\bibfield  {journal} {\bibinfo  {journal} {Rev. Mod. Phys.}\
  }\textbf {\bibinfo {volume} {78}},\ \bibinfo {pages} {217} (\bibinfo {year}
  {2006})}\BibitemShut {NoStop}%
\bibitem [{\citenamefont {Korotkov}\ \emph {et~al.}(1994)\citenamefont
  {Korotkov}, \citenamefont {Samuelsen},\ and\ \citenamefont
  {Vasenko}}]{Vasenko1994}%
  \BibitemOpen
  \bibfield  {author} {\bibinfo {author} {\bibfnamefont {A.~N.}\ \bibnamefont
  {Korotkov}}, \bibinfo {author} {\bibfnamefont {M.~R.}\ \bibnamefont
  {Samuelsen}}, \ and\ \bibinfo {author} {\bibfnamefont {S.~A.}\ \bibnamefont
  {Vasenko}},\ }\href@noop {} {\bibfield  {journal} {\bibinfo  {journal} {J.
  Appl. Phys.}\ }\textbf {\bibinfo {volume} {76}},\ \bibinfo {pages} {3623}
  (\bibinfo {year} {1994})}\BibitemShut {NoStop}%
\bibitem [{\citenamefont {Saira}\ \emph {et~al.}(2007)\citenamefont {Saira},
  \citenamefont {Meschke}, \citenamefont {Giazotto}, \citenamefont {Savin},
  \citenamefont {Mottonen},\ and\ \citenamefont {Pekola}}]{Pekola2007}%
  \BibitemOpen
  \bibfield  {author} {\bibinfo {author} {\bibfnamefont {O.-P.}\ \bibnamefont
  {Saira}}, \bibinfo {author} {\bibfnamefont {M.}~\bibnamefont {Meschke}},
  \bibinfo {author} {\bibfnamefont {F.}~\bibnamefont {Giazotto}}, \bibinfo
  {author} {\bibfnamefont {A.~M.}\ \bibnamefont {Savin}}, \bibinfo {author}
  {\bibfnamefont {M.}~\bibnamefont {Mottonen}}, \ and\ \bibinfo {author}
  {\bibfnamefont {J.~P.}\ \bibnamefont {Pekola}},\ }\href@noop {} {\bibfield
  {journal} {\bibinfo  {journal} {Phys. Rev. Lett.}\ }\textbf {\bibinfo
  {volume} {99}},\ \bibinfo {pages} {027203} (\bibinfo {year}
  {2007})}\BibitemShut {NoStop}%
\bibitem [{\citenamefont {Lefevre}\ \emph {et~al.}(2006)\citenamefont
  {Lefevre}, \citenamefont {Volz},\ and\ \citenamefont
  {Chapuis}}]{Chapuis2006}%
  \BibitemOpen
  \bibfield  {author} {\bibinfo {author} {\bibfnamefont {S.}~\bibnamefont
  {Lefevre}}, \bibinfo {author} {\bibfnamefont {S.}~\bibnamefont {Volz}}, \
  and\ \bibinfo {author} {\bibfnamefont {P.-O.}\ \bibnamefont {Chapuis}},\
  }\href@noop {} {\bibfield  {journal} {\bibinfo  {journal} {International
  Journal of Heat and Mass Transfer}\ }\textbf {\bibinfo {volume} {49}},\
  \bibinfo {pages} {251} (\bibinfo {year} {2006})}\BibitemShut {NoStop}%
\bibitem [{\citenamefont {Grosse}\ \emph {et~al.}(2011)\citenamefont {Grosse},
  \citenamefont {Bae}, \citenamefont {Lian}, \citenamefont {Pop},\ and\
  \citenamefont {King}}]{King2011}%
  \BibitemOpen
  \bibfield  {author} {\bibinfo {author} {\bibfnamefont {K.~L.}\ \bibnamefont
  {Grosse}}, \bibinfo {author} {\bibfnamefont {M.-H.}\ \bibnamefont {Bae}},
  \bibinfo {author} {\bibfnamefont {F.}~\bibnamefont {Lian}}, \bibinfo {author}
  {\bibfnamefont {E.}~\bibnamefont {Pop}}, \ and\ \bibinfo {author}
  {\bibfnamefont {W.~P.}\ \bibnamefont {King}},\ }\href@noop {} {\bibfield
  {journal} {\bibinfo  {journal} {Naure Nanotechnology}\ }\textbf {\bibinfo
  {volume} {6}},\ \bibinfo {pages} {287} (\bibinfo {year} {2011})}\BibitemShut
  {NoStop}%
\bibitem [{\citenamefont {Khurgin}(2007)}]{Khurgin2007}%
  \BibitemOpen
  \bibfield  {author} {\bibinfo {author} {\bibfnamefont {J.~B.}\ \bibnamefont
  {Khurgin}},\ }\href@noop {} {\bibfield  {journal} {\bibinfo  {journal} {Phys.
  Rev. Lett.}\ }\textbf {\bibinfo {volume} {98}},\ \bibinfo {pages} {177401}
  (\bibinfo {year} {2007})}\BibitemShut {NoStop}%
\bibitem [{\citenamefont {Liu}\ \emph {et~al.}(2011)\citenamefont {Liu},
  \citenamefont {Hsu},\ and\ \citenamefont {Chen}}]{Chen2011}%
  \BibitemOpen
  \bibfield  {author} {\bibinfo {author} {\bibfnamefont {Y.-S.}\ \bibnamefont
  {Liu}}, \bibinfo {author} {\bibfnamefont {B.~C.}\ \bibnamefont {Hsu}}, \ and\
  \bibinfo {author} {\bibfnamefont {Y.-C.}\ \bibnamefont {Chen}},\ }\href@noop
  {} {\bibfield  {journal} {\bibinfo  {journal} {J. Phys. Chem. C}\ }\textbf
  {\bibinfo {volume} {115}},\ \bibinfo {pages} {6111} (\bibinfo {year}
  {2011})}\BibitemShut {NoStop}%
\bibitem [{\citenamefont {Landau}\ and\ \citenamefont
  {Lifshitz}(1960)}]{landauVol8}%
  \BibitemOpen
  \bibfield  {author} {\bibinfo {author} {\bibfnamefont {L.~D.}\ \bibnamefont
  {Landau}}\ and\ \bibinfo {author} {\bibfnamefont {E.}~\bibnamefont
  {Lifshitz}},\ }\href@noop {} {\emph {\bibinfo {title} {Course of Theoretical
  Physics: Vol.: 8: Electrodynamics of Continuous Media}}}\ (\bibinfo
  {publisher} {Pergamon Press},\ \bibinfo {year} {1960})\BibitemShut {NoStop}%
\bibitem [{\citenamefont {Zwanzig}(1973)}]{Zwanzig1973}%
  \BibitemOpen
  \bibfield  {author} {\bibinfo {author} {\bibfnamefont {R.}~\bibnamefont
  {Zwanzig}},\ }\href@noop {} {\bibfield  {journal} {\bibinfo  {journal} {J.
  Stat. Phys.}\ }\textbf {\bibinfo {volume} {9}},\ \bibinfo {pages} {215}
  (\bibinfo {year} {1973})}\BibitemShut {NoStop}%
\bibitem [{\citenamefont {Bixon}\ and\ \citenamefont
  {Zwanzig}(1973)}]{Zwanzig1971}%
  \BibitemOpen
  \bibfield  {author} {\bibinfo {author} {\bibfnamefont {M.}~\bibnamefont
  {Bixon}}\ and\ \bibinfo {author} {\bibfnamefont {R.}~\bibnamefont
  {Zwanzig}},\ }\href@noop {} {\bibfield  {journal} {\bibinfo  {journal} {J.
  Stat. Phys.}\ }\textbf {\bibinfo {volume} {3}},\ \bibinfo {pages} {245}
  (\bibinfo {year} {1973})}\BibitemShut {NoStop}%
\bibitem [{\citenamefont {Devoret}\ \emph {et~al.}(1990)\citenamefont
  {Devoret}, \citenamefont {Esteve}, \citenamefont {Grabert}, \citenamefont
  {Ingold}, \citenamefont {Pothier},\ and\ \citenamefont
  {Urbina}}]{Urbina1990}%
  \BibitemOpen
  \bibfield  {author} {\bibinfo {author} {\bibfnamefont {M.~H.}\ \bibnamefont
  {Devoret}}, \bibinfo {author} {\bibfnamefont {D.}~\bibnamefont {Esteve}},
  \bibinfo {author} {\bibfnamefont {H.}~\bibnamefont {Grabert}}, \bibinfo
  {author} {\bibfnamefont {G.~L.}\ \bibnamefont {Ingold}}, \bibinfo {author}
  {\bibfnamefont {H.}~\bibnamefont {Pothier}}, \ and\ \bibinfo {author}
  {\bibfnamefont {C.}~\bibnamefont {Urbina}},\ }\href@noop {} {\bibfield
  {journal} {\bibinfo  {journal} {Phys. Rev. Lett.}\ }\textbf {\bibinfo
  {volume} {64}},\ \bibinfo {pages} {1824} (\bibinfo {year}
  {1990})}\BibitemShut {NoStop}%
\bibitem [{\citenamefont {Girvin}\ \emph {et~al.}(1990)\citenamefont {Girvin},
  \citenamefont {Glazman}, \citenamefont {Jonson}, \citenamefont {Penn},\ and\
  \citenamefont {Stiles}}]{Stiles1990}%
  \BibitemOpen
  \bibfield  {author} {\bibinfo {author} {\bibfnamefont {S.~M.}\ \bibnamefont
  {Girvin}}, \bibinfo {author} {\bibfnamefont {L.~I.}\ \bibnamefont {Glazman}},
  \bibinfo {author} {\bibfnamefont {M.}~\bibnamefont {Jonson}}, \bibinfo
  {author} {\bibfnamefont {D.~R.}\ \bibnamefont {Penn}}, \ and\ \bibinfo
  {author} {\bibfnamefont {M.~D.}\ \bibnamefont {Stiles}},\ }\href@noop {}
  {\bibfield  {journal} {\bibinfo  {journal} {Phys. Rev. Lett.}\ }\textbf
  {\bibinfo {volume} {64}},\ \bibinfo {pages} {3183} (\bibinfo {year}
  {1990})}\BibitemShut {NoStop}%
\bibitem [{\citenamefont {Devoret}\ and\ \citenamefont
  {Grabert}(1992)}]{devoret1992single}%
  \BibitemOpen
  \bibfield  {author} {\bibinfo {author} {\bibfnamefont {M.}~\bibnamefont
  {Devoret}}\ and\ \bibinfo {author} {\bibfnamefont {H.}~\bibnamefont
  {Grabert}},\ }\href@noop {} {\emph {\bibinfo {title} {Single Charge
  Tunneling}}},\ Vol.\ \bibinfo {volume} {264}\ (\bibinfo  {publisher} {New
  York, Plenum},\ \bibinfo {year} {1992})\BibitemShut {NoStop}%
\bibitem [{\citenamefont {Kobayashi}\ \emph {et~al.}(2012)\citenamefont
  {Kobayashi}, \citenamefont {Horiuchi}, \citenamefont {Kumai}, \citenamefont
  {Kagawa}, \citenamefont {Murakami},\ and\ \citenamefont
  {Tokura}}]{Tokura2012}%
  \BibitemOpen
  \bibfield  {author} {\bibinfo {author} {\bibfnamefont {K.}~\bibnamefont
  {Kobayashi}}, \bibinfo {author} {\bibfnamefont {S.}~\bibnamefont {Horiuchi}},
  \bibinfo {author} {\bibfnamefont {R.}~\bibnamefont {Kumai}}, \bibinfo
  {author} {\bibfnamefont {F.}~\bibnamefont {Kagawa}}, \bibinfo {author}
  {\bibfnamefont {Y.}~\bibnamefont {Murakami}}, \ and\ \bibinfo {author}
  {\bibfnamefont {Y.}~\bibnamefont {Tokura}},\ }\href@noop {} {\bibfield
  {journal} {\bibinfo  {journal} {Phys. Rev. Lett.}\ }\textbf {\bibinfo
  {volume} {108}},\ \bibinfo {pages} {237601} (\bibinfo {year}
  {2012})}\BibitemShut {NoStop}%
\bibitem [{\citenamefont {Miyamoto}\ \emph {et~al.}(2013)\citenamefont
  {Miyamoto}, \citenamefont {Yada}, \citenamefont {Yamakawa},\ and\
  \citenamefont {Okamoto}}]{Okamoto2013}%
  \BibitemOpen
  \bibfield  {author} {\bibinfo {author} {\bibfnamefont {T.}~\bibnamefont
  {Miyamoto}}, \bibinfo {author} {\bibfnamefont {H.}~\bibnamefont {Yada}},
  \bibinfo {author} {\bibfnamefont {H.}~\bibnamefont {Yamakawa}}, \ and\
  \bibinfo {author} {\bibfnamefont {H.}~\bibnamefont {Okamoto}},\ }\href@noop
  {} {\bibfield  {journal} {\bibinfo  {journal} {Nature Communications}\
  }\textbf {\bibinfo {volume} {4}},\ \bibinfo {pages} {2586} (\bibinfo {year}
  {2013})}\BibitemShut {NoStop}%
\bibitem [{\citenamefont {Fridkin}(2006)}]{Frid2006rev}%
  \BibitemOpen
  \bibfield  {author} {\bibinfo {author} {\bibfnamefont {V.~M.}\ \bibnamefont
  {Fridkin}},\ }\href@noop {} {\bibfield  {journal} {\bibinfo  {journal} {Phys.
  Usp.}\ }\textbf {\bibinfo {volume} {49}},\ \bibinfo {pages} {193} (\bibinfo
  {year} {2006})}\BibitemShut {NoStop}%
\bibitem [{\citenamefont {Fridkin}\ \emph {et~al.}(2010)\citenamefont
  {Fridkin}, \citenamefont {Gaynutdinov},\ and\ \citenamefont
  {Ducharme}}]{Frid2010rev}%
  \BibitemOpen
  \bibfield  {author} {\bibinfo {author} {\bibfnamefont {V.~M.}\ \bibnamefont
  {Fridkin}}, \bibinfo {author} {\bibfnamefont {R.~V.}\ \bibnamefont
  {Gaynutdinov}}, \ and\ \bibinfo {author} {\bibfnamefont {S.}~\bibnamefont
  {Ducharme}},\ }\href@noop {} {\bibfield  {journal} {\bibinfo  {journal}
  {Phys. Usp.}\ }\textbf {\bibinfo {volume} {53}},\ \bibinfo {pages} {199}
  (\bibinfo {year} {2010})}\BibitemShut {NoStop}%
\bibitem [{\citenamefont {Buron-LeCointe}\ \emph {et~al.}(2006)\citenamefont
  {Buron-LeCointe}, \citenamefont {Lemee-Cailleau}, \citenamefont {Cailleau},
  \citenamefont {Ravy}, \citenamefont {Berar}, \citenamefont {Rouziere},
  \citenamefont {Elkaim}, ,\ and\ \citenamefont {Collet}}]{Collet2006}%
  \BibitemOpen
  \bibfield  {author} {\bibinfo {author} {\bibfnamefont {M.}~\bibnamefont
  {Buron-LeCointe}}, \bibinfo {author} {\bibfnamefont {M.~H.}\ \bibnamefont
  {Lemee-Cailleau}}, \bibinfo {author} {\bibfnamefont {H.}~\bibnamefont
  {Cailleau}}, \bibinfo {author} {\bibfnamefont {S.}~\bibnamefont {Ravy}},
  \bibinfo {author} {\bibfnamefont {J.~F.}\ \bibnamefont {Berar}}, \bibinfo
  {author} {\bibfnamefont {S.}~\bibnamefont {Rouziere}}, \bibinfo {author}
  {\bibfnamefont {E.}~\bibnamefont {Elkaim}}, , \ and\ \bibinfo {author}
  {\bibfnamefont {E.}~\bibnamefont {Collet}},\ }\href@noop {} {\bibfield
  {journal} {\bibinfo  {journal} {Phys. Rev. Lett.}\ }\textbf {\bibinfo
  {volume} {96}},\ \bibinfo {pages} {205503} (\bibinfo {year}
  {2006})}\BibitemShut {NoStop}%
\bibitem [{\citenamefont {Huth}\ \emph {et~al.}(2014)\citenamefont {Huth},
  \citenamefont {Rippert}, \citenamefont {Sachser},\ and\ \citenamefont
  {Keller}}]{Plank2014}%
  \BibitemOpen
  \bibfield  {author} {\bibinfo {author} {\bibfnamefont {M.}~\bibnamefont
  {Huth}}, \bibinfo {author} {\bibfnamefont {A.}~\bibnamefont {Rippert}},
  \bibinfo {author} {\bibfnamefont {R.}~\bibnamefont {Sachser}}, \ and\
  \bibinfo {author} {\bibfnamefont {L.}~\bibnamefont {Keller}},\ }\href@noop {}
  {\bibfield  {journal} {\bibinfo  {journal} {Materials Research Express}\
  }\textbf {\bibinfo {volume} {1}},\ \bibinfo {pages} {046303} (\bibinfo {year}
  {2014})}\BibitemShut {NoStop}%
\bibitem [{\citenamefont {Grigalaitis}\ \emph {et~al.}(2010)\citenamefont
  {Grigalaitis}, \citenamefont {Banys}, \citenamefont {Macutkevic},
  \citenamefont {Adomavicius}, \citenamefont {Krotkus}, \citenamefont
  {Bormanis},\ and\ \citenamefont {Sternberg}}]{Sternberg2010}%
  \BibitemOpen
  \bibfield  {author} {\bibinfo {author} {\bibfnamefont {R.}~\bibnamefont
  {Grigalaitis}}, \bibinfo {author} {\bibfnamefont {J.}~\bibnamefont {Banys}},
  \bibinfo {author} {\bibfnamefont {J.}~\bibnamefont {Macutkevic}}, \bibinfo
  {author} {\bibfnamefont {R.}~\bibnamefont {Adomavicius}}, \bibinfo {author}
  {\bibfnamefont {A.}~\bibnamefont {Krotkus}}, \bibinfo {author} {\bibfnamefont
  {K.}~\bibnamefont {Bormanis}}, \ and\ \bibinfo {author} {\bibfnamefont
  {A.}~\bibnamefont {Sternberg}},\ }\href@noop {} {\bibfield  {journal}
  {\bibinfo  {journal} {Journal of the European Ceramic Society}\ }\textbf
  {\bibinfo {volume} {30}},\ \bibinfo {pages} {613} (\bibinfo {year}
  {2010})}\BibitemShut {NoStop}%
\end{thebibliography}%

\end{document}